\documentclass[journal]{IEEEtran}
\usepackage{amsfonts}
\usepackage{bm}
\usepackage{threeparttable}
\usepackage{graphicx}

\usepackage{amsmath}
\usepackage{multirow}
\usepackage{algorithm}
\usepackage{algorithmicx}
\usepackage{algpseudocode}
\usepackage{lipsum}
\usepackage{multicol}
\usepackage{float}
\usepackage{color}
\usepackage{url}
\usepackage{rotating}
\newtheorem{defn}{Definition}
\newtheorem{prob}{Problem}
\usepackage{graphics}
\usepackage{epsfig}

\usepackage{epstopdf}
\usepackage{algorithm}  
\usepackage{algpseudocode}  
\usepackage{amsmath}  

\ifCLASSINFOpdf
\else
\fi

\begin{document}


%


\title{A Survey of Distance-Based Vessel Trajectory Clustering: Data Pre-processing, Methodologies, Applications, and Experimental Evaluation}
\author{Maohan Liang,
	Ryan Wen Liu,~\IEEEmembership{Member,~IEEE,}
	Ruobin Gao,
	Zhe Xiao,
	Xiaocai Zhang,
	and Hua Wang 
	\thanks{ \textit{Corresponding author: Ruobin Gao}.}
	\thanks{M. Liang is with the Department of Civil and Environmental Engineering, National University of Singapore, Singapore (e-mail: mhliang@nus.edu.sg).}
	\thanks{R. W. Liu is with the Hubei Key Laboratory of Inland Shipping Technology, School of Navigation, Wuhan University of Technology, Wuhan, China (e-mail: wenliu@whut.edu.cn).}
	\thanks{R. Gao is with the School of Civil and Environmental Engineering, Nanyang Technological University, Singapore (e-mail: GAOR0009@e.ntu.edu.sg).}
	\thanks{Z. Xiao is with Institute of High Performance Computing, A*STAR, Singapore (e-mails: xiaoz@ihpc.a-star.edu.sg).}
	\thanks{X. Zhang is with the Department of Infrastructure Engineering, Faculty of Engineering and Information Technology, University of Melbourne, Victoria, Australia (e-mails: xiaocai.zhang@unimelb.edu.au).}
	\thanks{H. Wang is with the School of Automotive and Transportation Engineering, Hefei University of Technology, Hefei, China (e-mail: hwang191901@hfut.edu.cn).}
}

%
%

\markboth{
}%
{Liang \MakeLowercase{\textit{et al.}}: Bare Demo of IEEEtran.cls for IEEE Journals}
%
%
%
%
%
%
\maketitle

\begin{abstract}
	
	Vessel trajectory clustering, a crucial component of the maritime intelligent transportation systems, provides valuable insights for applications such as anomaly detection and trajectory prediction. This paper presents a comprehensive survey of the most prevalent distance-based vessel trajectory clustering methods, which encompass two main steps: trajectory similarity measurement and clustering. Initially, we conducted a thorough literature review using relevant keywords to gather and summarize pertinent research papers and datasets. Then, this paper discussed the principal methods of data pre-processing that prepare data for further analysis. The survey progresses to detail the leading algorithms for measuring vessel trajectory similarity and the main clustering techniques used in the field today. Furthermore, the various applications of trajectory clustering within the maritime context are explored. Finally, the paper evaluates the effectiveness of different algorithm combinations and pre-processing methods through experimental analysis, focusing on their impact on the performance of distance-based trajectory clustering algorithms. The experimental results demonstrate the effectiveness of various trajectory clustering algorithms and notably highlight the significant improvements that trajectory compression techniques contribute to the efficiency and accuracy of trajectory clustering. This comprehensive approach ensures a deep understanding of current capabilities and future directions in vessel trajectory clustering.
\end{abstract}

\begin{IEEEkeywords}
	Vessel trajectory, Trajectory clustering, Trajectory similarity, Intelligent transportation systems, Experimental evaluation
\end{IEEEkeywords}

%
\IEEEpeerreviewmaketitle


\section{Introduction}
\label{sec:Introduction}
\begin{figure*}[t]
	\centering
	\includegraphics[width=1\linewidth]{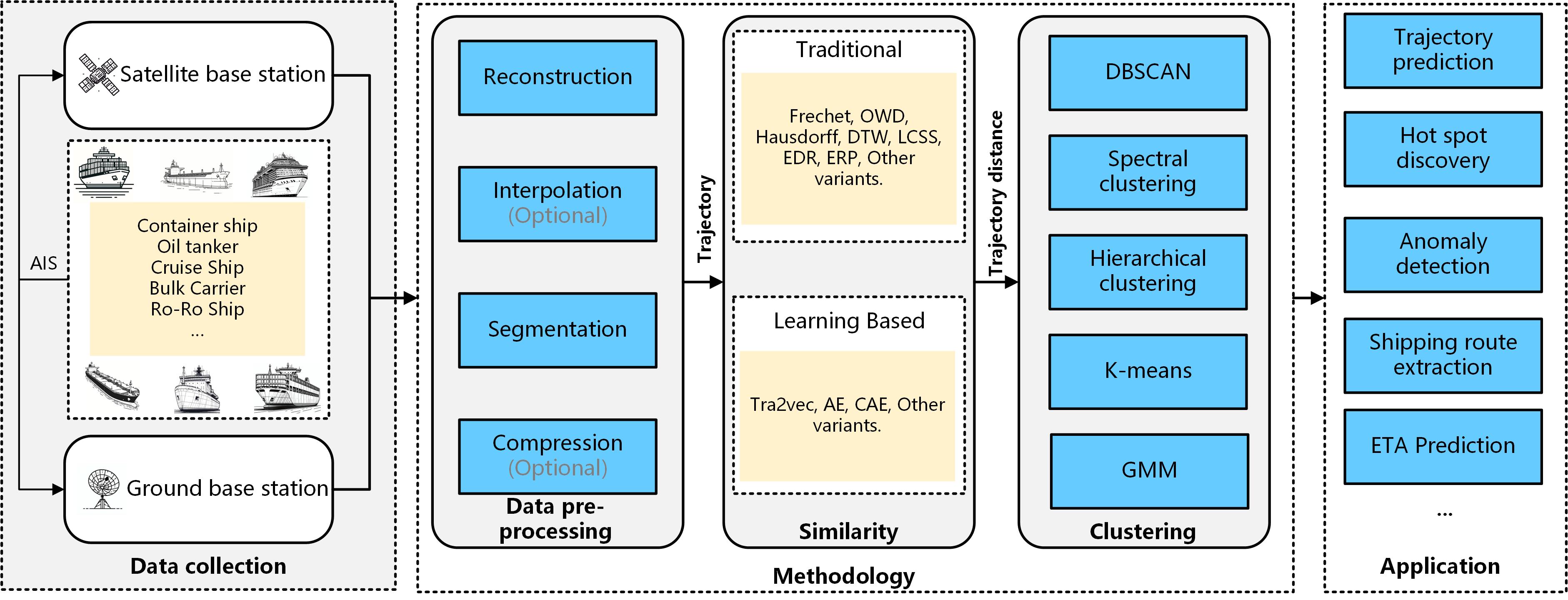}
	\caption{Paradigm of vessel trajectory clustering.}
	\label{fig:illumination}
\end{figure*}

%
\IEEEPARstart{V}{e}ssel trajectory clustering plays a vital role in maritime intelligent transportation systems (ITS) \cite{liang2022fine,li2022online,Meng2023,wang2022pathway}. It is potentially beneficial for maritime anomaly detection \cite{riveiro2018maritime}, vessel trajectory prediction \cite{hexeberg2017ais}, unmanned surface vehicles (USVs)  path planning \cite{liu2016unmanned}, etc. Given its growing importance and broad applicability \cite{shen2023analysis}, vessel trajectory clustering has become a hot topic in maritime research, highlighting the need for a detailed exploration of its recent advancements and developments. To address this need, we have conducted an extensive survey that thoroughly examines the various aspects of vessel trajectory clustering as depicted in the paradigm outlined in Fig. \ref{fig:illumination}. This survey aims to synthesize current knowledge and provide a cohesive overview of the field. 

The advancement of satellite positioning technology has significantly promoted vessel trajectory clustering. The automatic identification system (AIS) is a pivotal source for maritime data analysis \cite{ahmed2022spatio,wei2020self}. AIS is an automated tracking system used on vessels and by vessel traffic services for identifying and locating vessels by electronically exchanging data with other nearby vessels, AIS base stations, and satellites. AIS devices enable the global tracking of vessels in real time, generating extensive spatio-temporal data that outlines detailed vessel trajectories. These extensive data points are invaluable for the field of vessel trajectory data mining, particularly for vessel trajectory clustering \cite{huang2023maritime}.

Vessel trajectory clustering can be categorized into two main types \cite{yang2020tad,besse2016review}: distance-based clustering and density-based clustering. Density-based clustering \cite{yang2022maritime} groups trajectories based on the density of data points in the trajectory space. It identifies clusters as areas of high density separated by areas of low density, which is advantageous for discovering clusters of arbitrary shapes and for handling outliers effectively. Distance-based clustering \cite{yu2019trajectory} focuses on measuring the spatial distance between trajectories to group similar trajectories together. This method typically relies on geometric distances, aiming to minimize the variances within clusters while maximizing the differences between trajectories. It is especially useful for identifying groups of vessels following similar routes \cite{zhang2018data}. Therefore, the distance-based vessel trajectory clustering model is divided into two steps, i.e., similarity measurement and clustering algorithm. The focus of this paper is specifically on distance-based vessel trajectory clustering. 

The vessel trajectory clustering techniques play a crucial role in various maritime operations \cite{yang2019big}. For example, vessel trajectory clustering significantly enhances predictive models by improving the understanding of traffic patterns \cite{xiao2020big}. This advanced analysis allows predictive models to achieve greater accuracy in forecasting vessel movements \cite{feng2022collision}. Furthermore, identifying both typical and unusual traffic patterns improves maritime safety through early detection of potential hazards or illicit activities. Additionally, trajectory clustering is instrumental in managing traffic effectively in congested ports and contributes to environmental protection efforts by helping to ensure that vessels steer clear of ecologically sensitive areas \cite{rong2020data}. 

A large number of papers exist synthesizing methods, applications, and future directions for spatio-temporal trajectory similarity measurement \cite{hu2023spatio,yuan2014measuring,chen2023towards,yu2019trajectory}. 
However, these papers mainly focus on taxi trajectories and there are no current investigations on vessel trajectory clustering. Moreover, the current surveys mainly focus on trajectory similarity measurement, and few surveys experimentally evaluate distance-based vessel trajectory clustering. In conclusion, given the state-of-the-art research works, we conduct a comprehensive survey due to three following motivations:

\begin{itemize}
	\item Although a wide variety of methods have been proposed for vessel trajectory clustering, there currently exists a significant gap in the literature: a comprehensive review that systematically summarizes and synthesizes these diverse approaches is notably lacking. This absence of a consolidated overview makes it challenging for researchers and practitioners to fully understand the landscape of trajectory clustering techniques, compare their efficacy, and determine the most suitable methods for specific maritime applications. 
	\item A wide array of distance-based ship trajectory clustering methods have been developed, featuring numerous combinations of techniques and strategies. Given this variety, it is crucial to assess the effectiveness of these different combinations to understand their performance and applicability in various maritime situations. This evaluation is essential for determining which methods are most successful at accurately grouping similar vessel trajectories and for identifying how these methodologies can be optimized to enhance their utility in practical maritime operations. Such systematic assessments will help refine existing models and potentially drive innovation in trajectory clustering practices.
	\item Various data pre-processing methods are purported to enhance the effectiveness of vessel trajectory clustering, particularly for distance-based clustering approaches. It is important to critically assess these claims by evaluating how different pre-processing techniques influence the outcomes of distance-based vessel trajectory clustering. This assessment is crucial for identifying which pre-processing methods truly improve clustering accuracy and efficiency, thereby ensuring the most relevant and useful trajectories are analyzed. Understanding the impact of pre-processing can help in fine-tuning the clustering process, ultimately leading to more precise and reliable clustering results in maritime navigation and safety applications.
\end{itemize}

This paper is structured as follows: Section \ref{sec:Bib} provides a bibliometric analysis of the existing literature on trajectory clustering methods. Section \ref{sec:Pre} introduces key concepts, definitions, and problems relevant to our study.  Section \ref{sec:Data} focuses on data pre-processing techniques for vessel trajectory clustering, outlining methods to prepare data for subsequent analysis. Following this, Section \ref{sec:Distance} discusses Distance-based Vessel Trajectory Similarity Measurement techniques, exploring different metrics used to assess similarity between vessel trajectories. Building on the concepts from Section \ref{sec:Distance}, Section \ref{sec:Unsupervised} delves into various clustering algorithms that utilize these distance measurements. Section \ref{sec:Applications} examines the applications of Vessel Trajectory Clustering, demonstrating how these methods can be applied in real-world scenarios. Section \ref{sec:Experimental} presents an experimental evaluation of the discussed clustering algorithms, providing insights into their effectiveness and efficiency. Finally, the paper concludes with a summary in Section \ref{sec:Discussion}, where we summarize our findings and suggest directions for future research. This paper serves as an extension of the work originally presented in 5-th IEEE International Conference on Big Data Analytics \cite{lu2020shape}.

\section{Bibliometric Analysis}\label{sec:Bib}
To have a comprehensive overview of the current state of literature and research related to vessel trajectory clustering, this paper searched the literature from 2010-2023 through the Web of Science platform. We used `ship trajectory clustering' or `vessel trajectory clustering' as a keyword and retrieved papers that had the keyword in the abstract of the paper.
According to the search results of the website, there are 156 relevant papers in journals and conference proceedings as of December 2023. 

\begin{figure*}[t]
	\centering
	\includegraphics[width=0.8\linewidth]{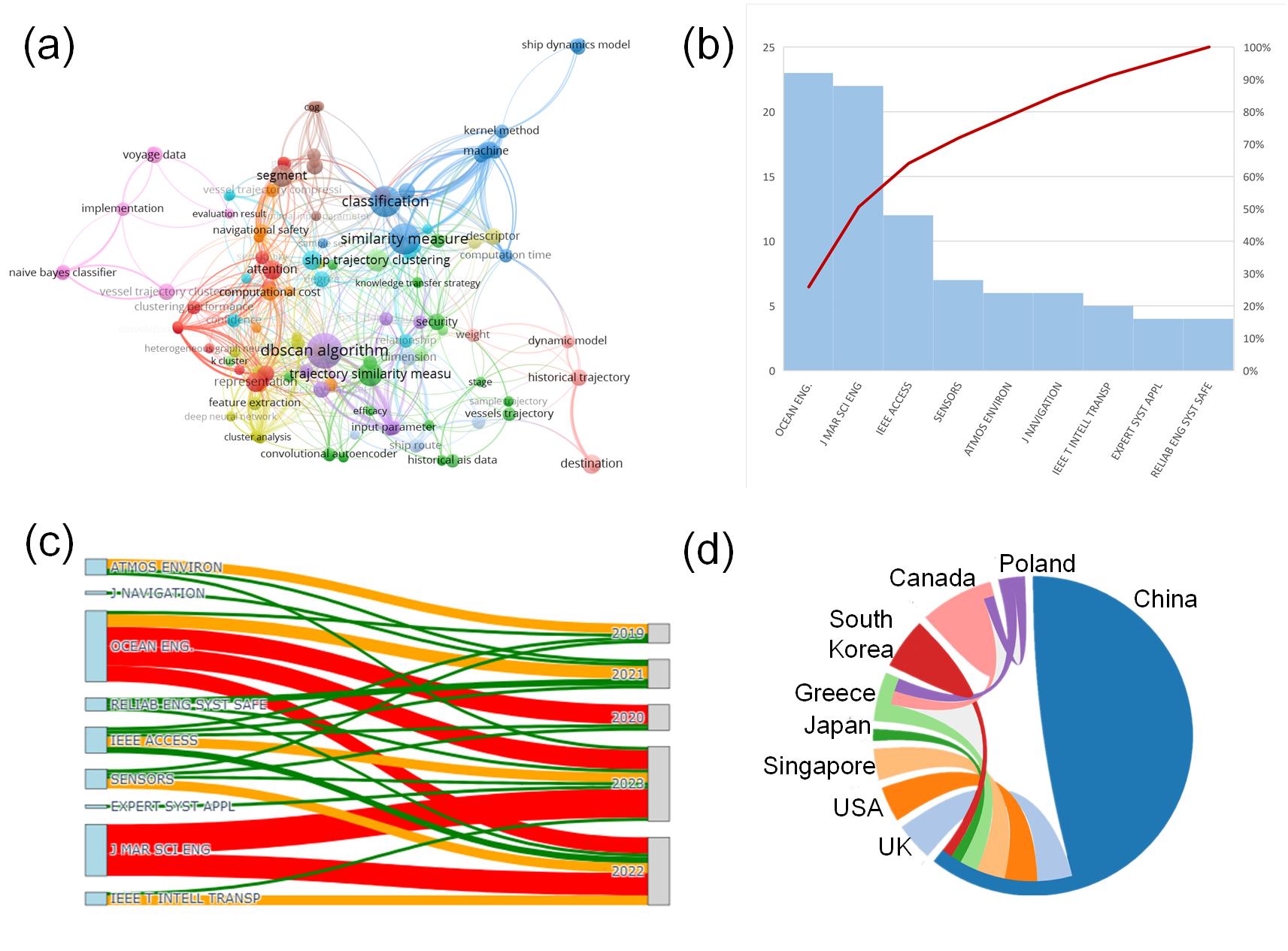}
	\caption{Bibliometric analysis of vessel trajectory clustering from 2010-2023.}
	\label{fig:literature}
\end{figure*}

Fig. \ref{fig:literature} (a) shows the results of literature keyword clustering according `CiteSpace', which is an tool for visualizing and analyzing scientific literature networks . It can be seen that vessel trajectory clustering contains a large number of traditional distance measures and deep learning methods. AIS is the main data source for vessel trajectory clustering. Application scenarios include destination port prediction, route extraction, navigation safety, etc.

As the Fig. \ref{fig:literature} (b) and (c) shown, these studies are mainly published in maritime research journals (e.g., Ocean Engineering, Journal of Marine Science and Engineering, Journal of Navigation), transportation research journals (e.g., IEEE Transactions on Intelligent Transportation Systems), safety research journals (e.g., Reliability Engineering \& System Safety), comprehensiveness journals (e.g., IEEE Access, Sensors), and several journals on artificial intelligence (e.g.,  Expert Systems with Applications). Fig. \ref{fig:literature} (b) depicts the time distribution of publication counts. From Fig. \ref{fig:literature} (c), we can
have an observation that there is a significant surge around the year 2023 in terms of the research work counts in this field. According to the country cooperation relationship graph in Fig. \ref{fig:literature} (d), we can see that Chinese authors are the main group of relevant papers published.

\section{Preliminaries}\label{sec:Pre}
In this section, we formally define the problem of trajectory similarity and trajectory clustering, Table \ref{tab:Notation} lists the frequently used notation in this paper.

\begin{table}[t]
	\centering
	\caption{Notation}
	\begin{tabular}{|l|l|}
		\hline
		$p_{i}$                & The $i$th point of vessel trajectory                                                                           \\ \hline	
		$T$                    & A set of trajectories                                                                                \\ \hline
		$T_{i}$                & The $i$th trajectory of set $T$                                                                           \\ \hline
		$\left \| L \right \|$ & Length of trajectory $L$                                                                                  \\ \hline
		$p_{T^{k}_{i}}$        & Point $k$ on trajectory at $T_{i}$                                                                        \\ \hline
		$p^{x},p^{y}$          & \begin{tabular}[c]{@{}l@{}}The Latitude and longitude \\ coordinates of trajectory point $p$\end{tabular} \\ \hline
		$dist(p_{i},p_{j})$    & \begin{tabular}[c]{@{}l@{}}The Euclidean distance between\\ $p_{i}$ and $p_{j}$\end{tabular}              \\ \hline
		$D(T_{i},T_{j})$    & \begin{tabular}[c]{@{}l@{}}Trajectory distance measure between\\ $T_{i}$ and $T_{j}$\end{tabular}              \\ \hline
	\end{tabular}\label{tab:Notation}
\end{table}

\subsection{Definition}
\begin{defn}
	\textbf{\emph{(Vessel trajectory)}} Vessel trajectory is a sequence of multi-dimensional data points sorted by timestamp, which is used to describe the spatial and temporal information of a vessel. Specifically, a vessel trajectory of length $n$ is defined as $T=\left \{ p_{1}, p_{2},... ,p_{n} \right \}$, where $p_{i}(1\leq i\leq n)$ represents the $i$-th sample point of the trajectory. Each sample point $p_{i}$ contains several pieces of information, including the longitude, latitude, speed, and heading of the vessel at that particular point in time. 
\end{defn}

\subsection{Problems}
\begin{prob}
	\textbf{\emph{(Trajectory similarity)}}  Trajectory similarity (a.k.a trajectory distance) measure is a defined metric used to evaluate how closely two trajectories resemble each other.  
\end{prob}
Trajectory similarity measure must adhere to the following properties:
\begin{enumerate}
	\item[1).] The distance (similarity measure) between two trajectories $D(T_{A}, T_{B})$ is zero if and only if $T_{A}$ and $T_{B}$ are exactly the same trajectory. This means that the only way for two trajectories to have the lowest possible distance, which is zero, is when they are identical.
	\item[2).] For any two trajectories $T_{A}$ and $T_{B}$, the distance $D(T_{A}, T_{B})$ is always greater than or equal to zero. This implies that the similarity measure is always a non-negative value, reflecting the fact that the dissimilarity between two trajectories cannot be a negative quantity. 
	\item[3).]For any two trajectories $T_{A}$ and $T_{B}$, the distance from $T_{A}$ to $T_{B}$ is the same as the distance from $T_{B}$ to $T_{A}$, formally stated as $D(T_{A}, T_{B})=D(T_{B}, T_{A})$. This indicates that the order in which trajectories are compared does not influence their distance or similarity measure. 
	\item[4).]For any three trajectories $T_{A}, T_{B}$, and $T_{C}$, the direct distance from $T_{A}$ to $T_{C}$ is always less than or equal to the sum of the distances from $T_{A}$ to $T_{B}$ and from $T_{B}$ to $T_{C}$, represented as $D(T_{A}, T_{C}) \leq D(T_{A}, T_{B})+D(T_{B}, T_{C})$. This property ensures that the ``shortcut" between two trajectories cannot be shorter than the "path" going through an intermediate trajectory.
\end{enumerate}

\begin{prob}
	\textbf{\emph{(Trajectory clustering)}}  Trajectory clustering is the process of grouping a collection of trajectories based on their similarity using clustering algorithms. In trajectory clustering, the goal is to find a way to measure the similarity between trajectories and to organize those with high similarity into clusters. 
\end{prob}

In addition, this paper will explore two controversial topics of current research.

\begin{prob}
	\textbf{\emph{(Which combinations of similarity measurement and clustering models give the good results in vessel trajectory clustering?)}}  There are numerous combinations of trajectory similarity measures and clustering models, and determining which algorithms are effective for vessel trajectory similarity measurement tasks is a hot topic of discussion.
\end{prob}

\begin{prob}
	\textbf{\emph{(Does trajectory pre-processing affect the effectiveness of vessel trajectory similarity measures and clustering?)}}  
	There are multiple methods for trajectory pre-processing, and which of these pre-processing steps can achieve superior results for vessel trajectory similarity measurement and clustering remains a contentious topic.
\end{prob}

\section{Data pre-processing for vessel trajectory clustering}\label{sec:Data}
\begin{table*}[]
	\centering
	\caption{Summary of literature related to vessel trajectory pre-processing.}
	\begin{tabular}{|c|c|c|c|}
		\hline
		Trajectory  pre-processing tasks                   & Categories          & Methodologies                                                & Ref.    \\ \hline
		\multirow{3}{*}{Vessel trajectory reconstruction} & Linear methods      & CSI, Linear regression, lagrange interpolation, etc.          &\cite{guo2021improved} \cite{zhang2018novel} \cite{zaman2023interpolation} \cite{shi2019research}\cite{kontopoulos2021distributed}             \\ \cline{2-4} 
		& Filtering methods   & Kalman   Filtering, Particle Filtering, etc.                 &   \cite{xiao2020big}\cite{gou2021fdia} \cite{zhang2020vessel}  \cite{karlgaard2013ares}       \\ \cline{2-4} 
		& Learning methods    & LSTM,  RNN, CNN, VRNN, GRU, etc.                            &    \cite{han2023interaction}\cite{liang2019neural} \cite{li2020ais}         \\ \hline
		\multirow{4}{*}{Vessel trajectory segmentation}   & Time-interval       & Time-interval   statistic.                                   &   \cite{fu2017finding} \cite{li2022visual} \cite{capobianco2021deep} \cite{varlamis2021building}         \\ \cline{2-4} 
		& Stop-detection      & DBSCAN,   KNN, etc.                                          &   \cite{lee2021ais}\cite{liu2022hybrid}\cite{kim2023way}\cite{wu2020identifying}          \\ \cline{2-4} 
		& Multi-criteria      & Multi-factor statistics such as speed, time, etc.            &    \cite{varlamis2021building} \cite{zygouras2021detecting}        \\ \cline{2-4} 
		& Semantic trajectory & Bayesian deep learning, IMM, etc.                            &   \cite{herrero2019ais}\cite{wu2022semantic}\cite{markos2021capturing}          \\ \hline
		\multirow{2}{*}{Vessel trajectory compression}    & Offline compression & DP,  TD-TR, Speed-Based (SP), Heading-Based (HD), etc.       &   \cite{zhao2019trajectory} \cite{huang2020gpu} \cite{zhang2016ais} \cite{li2022unsupervised}      \\ \cline{2-4} 
		& Online compression  & Dead-Reckoning   (DR), Opening-Window Time-Ratio (OWT), etc. &   \cite{sun2020vessel} \cite{yan2022development}  \cite{klein2020dead}        \\ \hline
		\multirow{2}{*}{Vessel trajectory interpolation}  & Linear  methods     & CSI, Linear regression, etc.        &     \cite{guo2021anomaly}        \\ \cline{2-4} 
		& Learning  methods   & LSTM,   RNN, CNN, VRNN, GRU, etc.                            &    \cite{liu2020data}\cite{yuan2020multi}         \\ \hline
	\end{tabular}\label{tab:prepro}
\end{table*}
Due to the influence of the collection equipment, the collection environment or human factors, the trajectory data generally has problems such as noise, data missing, and sampling different frequencies, which will cause interference to subsequent trajectory analysis. These problems may affect the effect of trajectory clustering. Therefore, effective processing of AIS data is necessary. This paper mainly introduces four methods of AIS trajectory data processing: trajectory reconstruction, trajectory segmentation, trajectory compression and trajectory interpolation. Table \ref{tab:prepro} summarizes the related literature and methodologies.

\subsection{AIS data source}\label{sec:AIS}
\begin{table*}[]
	\centering
	\caption{AIS datasource website.}
	\begin{tabular}{|c|c|c|c|c|}
		\hline
		Name             & Source/Publisher                         & Region                      & Cost                                     & Tags                  \\ \hline
		AISHub\cite{depellegrin2020effects}           & Crowdsource                              & Global Coastal Coverage Map & Free                                     & open              \\ \hline
		MarineCadastre\cite{meyers2022some}   & NOAA Bureau of Ocean   Energy Management & US Coastal                  & Free                                     & USA    \\ \hline
		HELCOM\cite{lensu2019big}           & Baltic Marine Environment                & Baltic Sea                  & Free                                     & Europe \\ \hline
		CruiseMapper\cite{hoffmann2020plague}     & CruiseMapper                             & World                       & Free                                     & Cruise            \\ \hline
		Marine Traffic\cite{le2018can}    & Marine Traffic                           & Global                      & Credit & private           \\ \hline
		ExactAIS Archive\cite{ivanov2016distribution} & ExactAIS                                 & Global                      & Credit or Subscription                   & private           \\ \hline
		VesselFinder\cite{rao2021predicting}     & VesselFinder Ltd.                        & Global Coastal              & Credit or Subscription & private           \\ \hline
		FleetMon\cite{le2018can}         & FleetMon                                 & Global                      & Credit or Subscription                   & private           \\ \hline
	\end{tabular}\label{tab:AIS}
\end{table*}
As previously introduced, AIS data is widely used as the primary data source for vessel trajectory data mining. For the sake of scientific research, numerous government departments and companies have opened up access to and use of AIS data. Table \ref{tab:AIS} shows some of the AIS datasets. To the best of our limited knowledge, there is no completely free global online source of AIS data. This is mainly due to the high cost of AIS data maintenance and collection. AISHub \cite{depellegrin2020effects}  provides a crowd-sourced AIS data exchange platform. Users can provide their own online data in exchange for other online data on the platform. In addition, the United States \cite{meyers2022some} and Australia \cite{goldsworthy2017spatial} have open-sourced their coastal offline AIS data, which has been widely used in scientific research.

\subsection{Vessel trajectory reconstruction}
Trajectory noise and missing are unavoidable due to many factors such as sensor failure, detection technology errors, interference signals, or different sampling rates \cite{zhang2018novel}. The presence of noise may cause the effective information in the data to be overwhelmed, affecting the subsequent analysis and research of the data \cite{wang2020clustering,yi2018trajectory}. In general, trajectory reconstruction methods can be categorized into linear, filtering and learning methods. Linear methods are used to repair missing and noisy by constructing a regression function to fit the motion curve. The commonly used linear methods include spherical linear interpolation \cite{huang2020gpu,zhang2016ais}, Lagrange interpolation \cite{kontopoulos2021distributed}, Bezier curves \cite{li2019toward}, etc. Filters are popular methods used for signal processing, particularly suited for extracting useful information from noisy data \cite{wei2020particle}. Their key feature is the ability to estimate the state of a system from imperfect observational data, mainly by reducing the impact of uncertainty and noise. Filtering methods used in trajectory reconstruction include the Kalman Filter \cite{song2020automated}, extended Kalman Filter \cite{gou2021fdia}, Particle filter \cite{ristic2008statistical}, moving average filter \cite{abebe2022ship}, etc \cite{wang2019extraction}. Recently, with the development of deep learning techniques, the learning model demonstrated powerful results on the trajectory reconstruction task \cite{han2023interaction}. Learning-based methods include techniques like Recurrent Neural Networks (RNNs) \cite{han2023interaction}, Long Short-Term Memory networks (LSTMs) \cite{liang2019neural}, and Gated Recurrent Units (GRUs)  \cite{li2020ais}.

\subsection{Vessel trajectory segmentation}
Segmentation is the initial and critical step in spatio-temporal data analysis. Trajectory segmentation can be categorized as time-interval-based trajectory segmentation, stop-detection trajectory segmentation, multi-criteria trajectory segmentation, and semantic trajectory segmentation. Time-interval-based segmentation models segment the trajectories based on time intervals between two approximate data records \cite{zaman2023interpolation}. Similarly, latitude or speed difference also can be used to segment vessel trajectory \cite{wang2021ais}. To avoid the undesirable effects of missing data, stop-detection approaches are widely applied in vessel trajectory segmentation \cite{eljabu2022spatial}. These approaches used port as the segmentation point to segment vessel trajectory as sub-trajectory \cite{lee2021ais,liu2022hybrid,kim2023way,wu2020identifying}. Multi-criteria segmentation models are frequently employed to address more complex tasks and applications \cite{varlamis2021building}. These models segment vessel trajectories using a variety of methods such as time intervals and stopping points to minimize false alarms caused by segmentation errors \cite{zygouras2021detecting}. Semantic trajectory segmentation approaches have been widely studied in road traffic. They are often used to identify semantic trajectories such as a set of trips and activities that are of interest for a given application, and partitions them based on semantic points, such as stay points or transportation transit points. For example, Markos et al. \cite{markos2021capturing} introduced an unsupervised GPS trajectory segmentation method based on Bayesian deep learning, aiming to identify and classify different transportation modes. In maritime research, Wu et al. \cite{wu2022semantic} developed a window-based segmentation algorithm to segment vessel trajectory and classifier trajectory as fishing or non-fishing. Similarly, Herrero et al. \cite{herrero2019ais} introduced a system using the Interacting Multiple Model (IMM) filter to segment vessel trajectories for classifying vessel trajectories by types and activities.

\subsection{Vessel trajectory compression}
Trajectory data compression, also known as (a.k.a) trajectory data downsampling \cite{zhu2021ship}. The trajectory data often has the problem of trajectory point redundancy. An over abundance of vessel trajectories may cause a significant reduction in the efficiency of the algorithmic analysis \cite{zhao2018method}. Therefore, extracting the core features from over-abundant information trajectories is an important topic. Vessel trajectory compression methods are generally divided into two categories: offline compression and online compression \cite{wei2020ais}. These two types of compression methods can satisfy different application scenarios and requirements. Specifically, online compression algorithms are adept at processing data streams and minimizing data storage requirements. In contrast, offline compression algorithms enhance the performance of other algorithms by optimizing data beforehand. Notable offline compression techniques include the Douglas-Peucker algorithm \cite{fikioris2023optimizing}, Theo-Pavlidis (TP) algorithm \cite{lin2021error}, and synchronous Euclidean distance algorithm \cite{lin2019one}. On the other hand, online compression methods encompass algorithms such as dead reckoning \cite{klein2020dead}, open window algorithm \cite{tang2021novel}, vector-based online trajectory compression algorithm \cite{cai2023voltcom}.

\subsection{Vessel trajectory interpolation}
Trajectory data interpolation, a.k.a trajectory data upsampling \cite{zhao2024hitrip,du2022lifelong}.
It aimed at synchronizing timing across various trajectories or filling in missing segments, significantly boosts the robustness of subsequent applications. This process is crucial for applications requiring consistent and complete route information, such as navigation systems, tracking applications, and movement analysis, where gaps or asynchronous data can lead to inaccuracies or inefficiencies. Vessel trajectory interpolation also can be categorized into linear and learning-based methods. Linear approaches encompass techniques such as cubic spline interpolation (CSI) and kinematic fitting, etc \cite{sun2024multi}. On the other hand, learning-based methods include techniques like RNNs, LSTMs, and GRUs \cite{liu2020data,yuan2020multi}. For example, Zhao et al. \cite{zhao2024hitrip} introduced a deep learning approach that interpolates historical vessel monitoring system (VMS) data from two-hour intervals down to three minutes by harnessing both VMS and marine hydrological datasets. Zaman  et al. \cite{zaman2023interpolation} presented a two-step approach utilizing interpolation method to effectively detect waypoints in vessel trajectories from sparse AIS data. Guo et al. \cite{guo2021anomaly} introduced a method for detecting anomalies in AIS trajectory data through a three-step process involving data pre-processing, kinematic estimation, and error clustering, significantly enhancing the reliability and accuracy of vessel trajectory information extracted from raw AIS data.

\section{Distance-based Vessel trajectory similarity measurement}\label{sec:Distance}
According to the methods and processes, trajectory similarity measurement algorithms can be categorized in various ways. The classification of trajectory similarity measurement algorithms can be shown in the table. Chang et al. \cite{chang2023trajectory} categorized the trajectory similarity metric algorithms into Heuristic and Learned. Similarly, Hu et al. \cite{Hu2023} categorized trajectory similarity metric algorithms into learned and unlearned. The differences in these classifications are mainly centered on the unlearned trajectory similarity measurement algorithms. Since the cars are limited by the road network, the cars trajectory similarity can be calculated through the nodes of the road network. Theoretically, vessel routes are infinite, thus vessel trajectory similarity measurement algorithms seldom consider shipping network limitations. Given these considerations, this paper classifies vessel trajectory similarity measurement into traditional methods and learning-based methods. The traditional methods and learning methods are described separately below.

\begin{table}[]
	\centering
	\caption{Summary of literature related to Vessel trajectory similarity measurement.}
	\begin{tabular}{|c|c|}
		\hline
		Methods   & Reference                                                                                                                                  \\ \hline
		DTW       & \cite{zhao2019novel,wei2020ais,bai2023adaptive,xu2019use,guo2023asynchronous,xie2024anomaly,liu2023shipping,huang2020gpu} \\ \hline
		LCSS      & \cite{widyantara2023automatic,park2021ship,le2013unsupervised,yoo2022statistical}                                        \\ \hline
		EDR     &  \cite{sheng2018research,nie2021trajectory}                                                                               \\ \hline
		ERP       & \cite{zhang2021trajectory}                                                                                          \\ \hline
		OWD       & \cite{ma2014vessel}                                                                                                  \\ \hline
		Fr{\'e}chet   & \cite{andersen2021ais,cao2018pca}                                                                                         \\ \hline
		Hausdorff & \cite{wang2021ship,yang2022maritime,arguedas2017maritime,hu2022intelligent,alessandrini2018estimated}                    \\ \hline
		Seq2seq   & \cite{ yao2017trajectory,yao2018learning,murray2021ais}                                                                   \\ \hline
		CAE       & \cite{wang2021ais,duan2022semi}           \\ \hline                                                                                
	\end{tabular}
\end{table}

\begin{figure*}[t]
	\centering
	\includegraphics[width=0.9\linewidth]{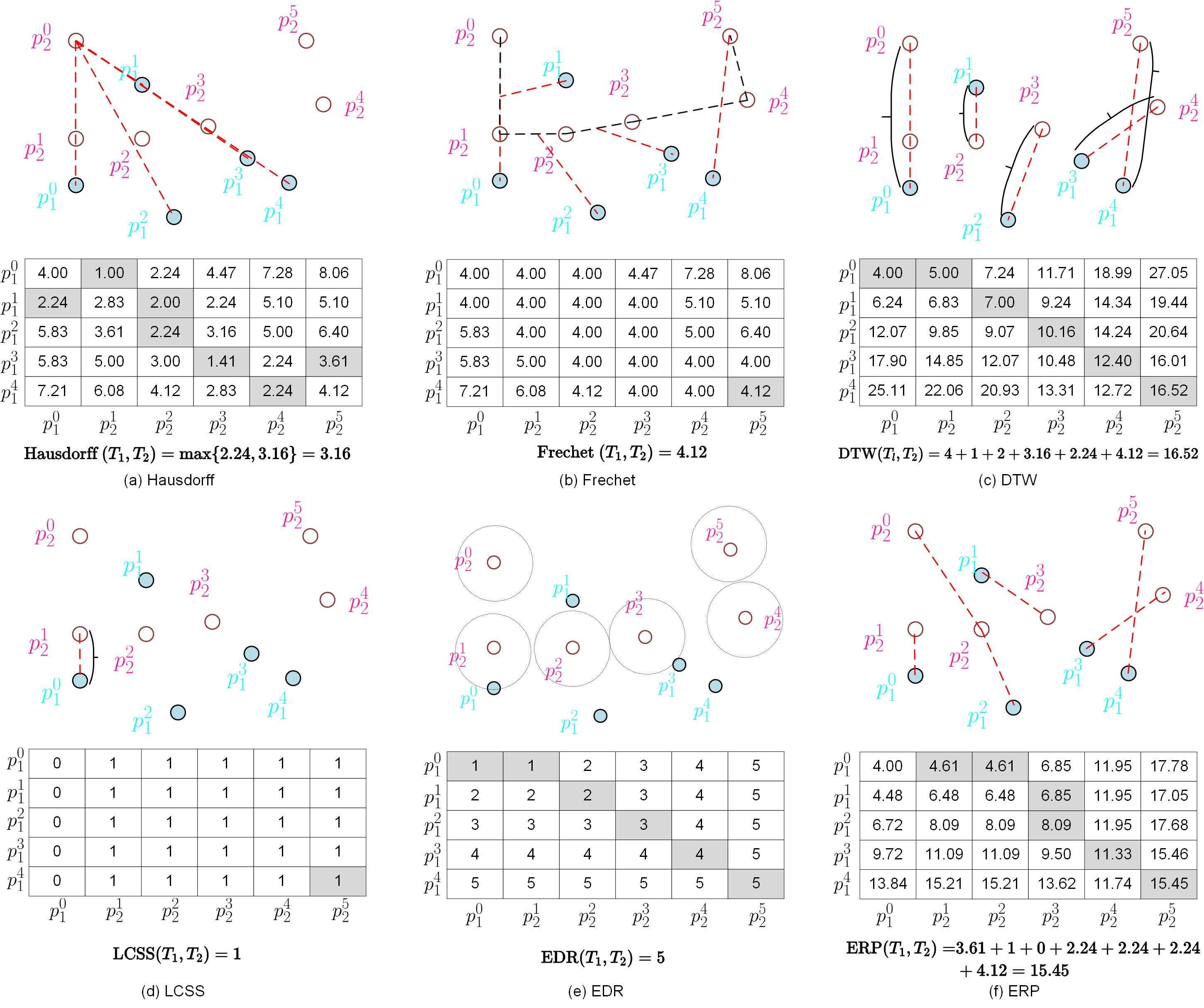}
	\caption{Illustration of the traditional trajectory similarity measurement \cite{hu2023spatio}.}
	\label{fig:similarity}
\end{figure*}

\subsection{Traditional trajectory similarity measurement methods}
Traditional trajectory similarity measurement methods are widely used with the task of calculating vessel trajectory similarity. Traditional vessel trajectory similarity measurement methods include Hausdorff distance, Fr{\'e}chet distance, One-Way Distance (OWD), Dynamic Time Warping (DTW), Longest Common Subsequence (LCSS), Edit Distance on Real Sequence (EDR) and Edit Distance with Real Penalty (ERP).

\subsubsection{Hausdorff distance}
The Hausdorff distance \cite{sousa2020vehicle} is particularly useful as it captures the maximum discrepancy between two trajectories. This measurement can accommodate inconsistencies due to variations in point density along the trajectories and is robust against minor perturbations. However, it is important to note that the Hausdorff distance may be sensitive to noise in the data. The Hausdorff distance measures the extent of the mismatch between two trajectories by calculating the greatest of all the distances from a point on one trajectory to the closest point on the other trajectory. Let us denote the Hausdorff distance as $D_{\text{hausdorff}}$, represented by the equation below:

\begin{equation}\label{eq:Hausdorff}
	D_{\text{hausdorff}}\left ( T_{i}, T_{j}\right ) = \max \left \{ h\left ( T_{i}, T_{j} \right ), h\left ( T_{j}, T_{i} \right ) \right \}
\end{equation}
where the one-way Hausdorff distances are defined as follows:
\begin{equation}
	h\left ( T_{i}, T_{j} \right ) = \max_{p \in T_i} \min_{q \in T_j} \text{dist}(p, q)
\end{equation}
where $h\left ( T_{i}, T_{j} \right )$ calculates the one-way Hausdorff distance from trajectory $T_i$ to $T_j$ by finding the maximum distance from any point $p$ on $T_i$ to its nearest point $q$ on $T_j$. Similarly, $h\left ( T_{j}, T_{i} \right )$ represents the one-way distance from $T_j$ to $T_i$. The bidirectional Hausdorff distance, $D_{\text{hausdorff}}\left ( T_{i}, T_{j}\right )$, is the maximum of these one-way distances and provides a comprehensive measure of the similarity between the two trajectories.

\subsubsection{Fr{\'e}chet distance}
The Fr{\'e}chet distance \cite{niu2019label} is a measure of similarity between two trajectories that takes into account the sequence and position of points along the paths. It is often described metaphorically as the "dog leash distance." To illustrate, consider a scenario where a dog and its owner are each walking along different paths; the Fr{\'e}chet distance represents the minimum length of a leash required for both to traverse their respective paths without parting ways. The Fr{\'e}chet distance uniquely captures the continuity and ordering of the points along the trajectories, making it highly sensitive to the paths' actual shapes. However, it is worth noting that since this metric considers the maximum of all point-to-point distances (under the optimal reparameterization), it can be sensitive to outliers or anomalies in the data. The distance is mathematically defined as follows:

\begin{equation}
	D_{\text{Fr{\'e}chet}}\left ( T_{i}, T_{j}\right ) = \min_{\alpha, \beta} \max_{t \in [0,1]} \text{dist}\left ( T_{i}(\alpha(t)), T_{j}(\beta(t)) \right )
\end{equation}
Here, $\alpha$ and $\beta$ are continuous and non-decreasing functions from $[0, 1]$ to the parameter spaces of the trajectories $T_i$ and $T_j$, which ensure that the points $T_i(\alpha(t))$ and $T_j(\beta(t))$ correspond to positions on the trajectories at "time" $t$. The functions $\alpha$ and $\beta$ effectively synchronize the movements along the trajectories. In this equation, $D_{\text{Fr{\'e}chet}}\left ( T_{i}, T_{j}\right )$ denotes the Fr{\'e}chet distance between trajectories $T_i$ and $T_j$, where the trajectories are of lengths $m$ and $n$, respectively.

\subsubsection{OWD distance}
The One Way Distance (OWD) \cite{besse2016review} framework provides a comprehensive measure that encompasses both the shape and physical distance of the trajectories, and is less sensitive to noise due to its averaging approach.
The OWD measures the directional discrepancy from one trajectory to another and is inherently asymmetric. However, the average of the OWD distances in both directions between two trajectories is symmetric, as expressed in the following equation:

\begin{equation}
	D_{\text{OWD}}\left ( T_{i}, T_{j}\right ) = \frac{1}{2}\left ( D_{o}\left ( T_{i}, T_{j} \right ) + D_{o}\left ( T_{j}, T_{i} \right ) \right )
\end{equation}
where the individual one-way distances are defined as:
\begin{equation}
	D_{o}\left ( T_{i}, T_{j} \right ) = \frac{1}{|L_{i}|} \sum_{p \in T_i} D_{\text{point}}(p, T_{j})
\end{equation}

\begin{equation}
	D_{o}\left ( T_{j}, T_{i} \right ) = \frac{1}{|L_{j}|} \sum_{p \in T_j} D_{\text{point}}(p, T_{i})
\end{equation}
where $D_{\text{OWD}}\left ( T_{i}, T_{j}\right )$ represents the symmetric OWD distance between trajectories $T_{i}$ and $T_{j}$. $D_{o}\left ( T_{i}, T_{j} \right )$ and $D_{o}\left ( T_{j}, T_{i} \right )$ denote the asymmetric one-way distances from $T_{i}$ to $T_{j}$ and $T_{j}$ to $T_{i}$, respectively. These distances are calculated by averaging the minimum distances from every point pp on one trajectory to the other trajectory, normalized by the total length of the trajectory segment from which the points are taken ($|L_i|$ or $|L_j|$).

The minimum point-to-trajectory distance, $D_{\text{point}}(p, T_i)$, is defined as the shortest distance from a point pp to any point on trajectory $T_i$:

\begin{equation}
	D_{\text{point}}(p, T_{i}) = \min_{q \in T_i} \text{dist}(p, q)
\end{equation}

\subsubsection{DTW distance}
Dynamic time warping (DTW) \cite{kumar2018fast} is to complete the local scaling of the time dimension by repeating the previous points, and use the minimum distance between the trajectories as the DTW distance. DTW distance can better find similar trajectories after local scaling in the time dimension, effectively solving the problem of inconsistency between different sampling rates and time scales. However, when calculating the DTW distance, the trajectory sampling points must be continuous, so the DTW distance is sensitive to noise. In addition, the DTW distance is not suitable for trajectory distance calculation with large differences in a small range. The calculation method of DTW distance is as follows:

\begin{scriptsize}
	\begin{flalign}
		& D_{DTW}(T_i, T_j) = & \nonumber \\
		& \begin{cases}
			0, & \text{if } m = n = 0; \\
			\infty, & \text{if } m = 0 \text{ or } n = 0; \\
			\text{dist}(p_{T_i^1}, p_{T_j^1}) + \min
			\begin{cases}
				D_{DTW}(\text{R}(T_i), \text{R}(T_j)), \\
				D_{DTW}(\text{R}(T_i), T_j), \\
				D_{DTW}(T_i, \text{R}(T_j)),
			\end{cases}
			& \text{otherwise.}
		\end{cases} &
	\end{flalign}
\end{scriptsize}
where $D_{DTW}\left ( T_{i}, T_{j}\right )$ is the DTW distance between the trajectories $T_{i}$ and $T_{j}$ of length $m$ and $n$.  $R(T_{i})$ and $R(T_{j})$ represent the trajectory segment after removing the first point. As shown in formula (7), when the length of two trajectories is 0, the DTW distance of the two is 0, and the length of one trajectory is 0, the DTW distance is $\propto$ . When the length of the two trajectories is not 0, the DTW distance is the minimum distance between the trajectories calculated by the recursive method.

\subsubsection{LCSS distance}

The Longest Common Subsequence (LCSS) \cite{ma2023potential} distance differs from other trajectory similarity metrics as it aims to identify the longest common subsequence between two trajectory sequences. This length is then used to assess the similarity between the trajectories. LCSS is particularly useful for comparing high-dimensional time series or spatial-temporal trajectories as it is robust against noise due to its reliance on a threshold-based similarity criterion. However, setting the optimal thresholds ($\sigma$ and $\varepsilon$) is challenging: too high a threshold might dilute meaningful distinctions by oversimplifying similarity, whereas too low a threshold might overlook significant alignments by being overly restrictive. Moreover, LCSS does not differentiate between trajectories that share the same subsequence length but differ in other aspects. The LCSS is typically calculated using a recursive approach, as:

\begin{scriptsize}
	\begin{flalign}
		& LCSS(T_{i}, T_{j}) = & \nonumber \\
		& \begin{cases}
			0, & \text{if } m = n = 0; \\
			1 + LCSS_{\sigma, \varepsilon}(\text{R}(T_{i}), \text{R}(T_{j})), & \text{if } |p_{T_{i}^k}^x - p_{T_{j}^k}^x| \leq \sigma \text{ and } |p_{T_{i}^k}^y - p_{T_{j}^k}^y| \leq \varepsilon; \\
			\max \begin{cases}
				LCSS_{\sigma, \varepsilon}(\text{R}(T_{i}), T_{j}), \\
				LCSS_{\sigma, \varepsilon}(\text{R}(T_{j}), T_{i}),
			\end{cases}
			& \text{otherwise.}
		\end{cases} &
	\end{flalign}
\end{scriptsize}

where $LCSS_{\sigma, \varepsilon}\left ( T_{i}, T_{j}\right )$ indicates the length of the longest common subsequence, under the conditions that the differences in the $x$ and $y$ coordinates of points from the two trajectories do not exceed the thresholds $\sigma$ and $\varepsilon$, respectively. The function $\text{R}(T_i)$ represents the trajectory excluding its first point.

To convert the LCSS into a measure of distance that quantifies dissimilarity, the following transformation is used:

\begin{equation}
	D_{LCSS}\left ( T_{i}, T_{j}\right ) = 1 - \frac{LCSS\left ( T_{i}, T_{j}\right )}{\min(m, n)}
\end{equation}

Here, $D_{LCSS}\left ( T_{i}, T_{j}\right )$ is the normalized LCSS distance, which inversely correlates with the raw LCSS length to produce a similarity index ranging from 0 to 1, where 0 indicates identical trajectories and 1 indicates no commonality.

\begin{figure*}[]
	\centering
	\includegraphics[width=0.9\linewidth]{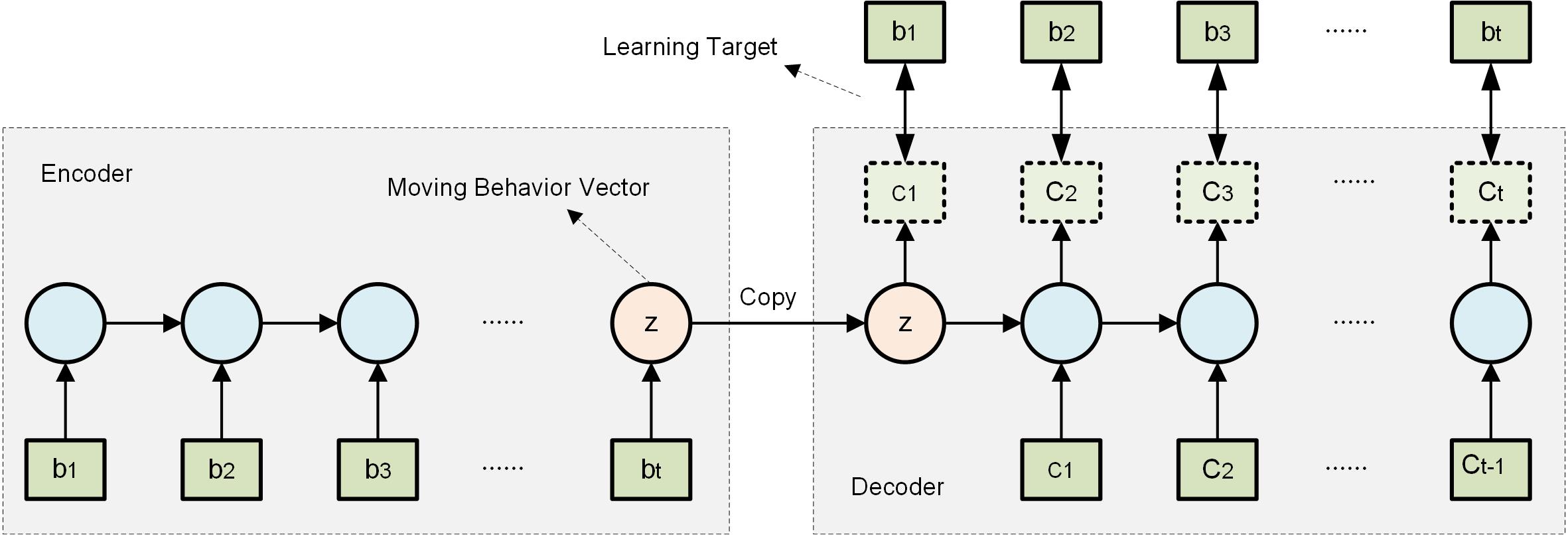}
	\caption{The structure of the sequence-to-sequence model \cite{yao2018learning}.}
	\label{fig:seq2seq}
\end{figure*}

\subsubsection{EDR distance}
The Edit Distance on Real sequence (EDR) \cite{tao2021comparative} measures the similarity between two sequences by counting the minimum number of editing operations required to transform one sequence into another. Originally used in bioinformatics and speech recognition, EDR is also applicable to numeric sequences such as trajectories. This metric is especially useful as it can be normalized to a value between 0 and 1. The EDR between two trajectories $T_i$ and $T_j$ is computed as follows:

\begin{small}
	\begin{flalign}
		& \operatorname{EDR}(T_i, T_j) = &  \nonumber\\
		& \begin{cases}
			m, & \text{if } n = 0; \\
			n, & \text{if } m = 0; \\
			\min \left\{
			\begin{array}{l}
				\operatorname{EDR}(\operatorname{R}(T_i), \operatorname{R}(T_j)) + \text{sub cost}, \\
				\operatorname{EDR}(\operatorname{R}(T_i), T_j) + 1, \\
				\operatorname{EDR}(T_i, \operatorname{R}(T_j)) + 1,
			\end{array}
			\right. & \text{otherwise.}
		\end{cases} &
	\end{flalign}
\end{small}
where the substitution cost is defined by the distance threshold $\varepsilon$:

\begin{equation}
	sub cost=\left\{\begin{matrix}
		0,\ \left | p_{t_{i}^{k}}^{x}- p_{t_{i}^{k}}^{x}\right |  \leq \varepsilon ,\ and \left | p_{t_{i}^{k}}^{y}- p_{t_{i}^{k}}^{y}\right |  \leq \varepsilon  \\1 ,\ \text { otherwise } 
		
	\end{matrix}\right.
\end{equation}

The normalized EDR distance is calculated as:

\begin{equation}
	D_{EDR} = \frac{EDR(T_{i}, T_{j})}{\max(m,n)}
\end{equation}

This normalization process maps the EDR value to a range between 0 and 1, where 0 indicates perfect similarity and 1 indicates maximum dissimilarity. The condition $|p_{T_{i}}^x - p_{T_{j}}^x| \leq \varepsilon$ and $|p_{T_{i}}^y - p_{T_{j}}^y| \leq \varepsilon$ checks whether the two points from $T_i$ and $T_j$ are considered to match under the set threshold. EDR is robust against noise, making it a suitable choice for applications where trajectories may include errors or irregularities. However, the accuracy of the EDR measurement significantly depends on the appropriateness of the threshold $\varepsilon$. As with LCSS, setting this parameter requires careful consideration to balance sensitivity and specificity. 

\subsubsection{ERP distance}
Edit Distance with Real Penalty (ERP) \cite{wang2021survey} integrates both Lp-norms and edit distance, enhancing the measurement of sequence similarity by taking into account local shifts and the treatment of gaps. Unlike classical edit distance, which uses a constant value for computing the cost of operations, ERP varies the cost depending on the real value difference between elements, except when a gap occurs. This approach enables the comparison of sequences even when there are variations in local values and allows for better handling of sequences with missing data.

The distance between two elements $a$ and $b$ in the context of ERP, considering the possibility of gaps, is defined as:
\begin{equation}
	D_a(a, b)= \begin{cases}|a-s|, & \text { if } b \text { is a gap } \\ |b-g|, & \text { if } a \text { is a gap } \\ |a-b|, & \text { otherwise }\end{cases}
\end{equation}
where $s$ is the start or end value of a trajectory used as a reference value for a gap in the other trajectory, and $g$ is a constant value for the gap penalty.

Given two trajectories $T_i$ and $T_j$, the ERP distance is computed by evaluating the cost of aligning each element from one trajectory to the other, including the introduction of gaps when necessary. This can be recursively expressed as follows:

\begin{footnotesize}
	\begin{flalign}
		& D_{ERP}(T_i, T_j) = & \nonumber \\
		& \begin{cases} 
			\sum_{k=1}^n |b_k - g|, & \text{if } m = 0; \\
			\sum_{k=1}^m |a_k - g|, & \text{if } n = 0; \\
			\min \left\{
			\begin{array}{l}
				D_{ERP}(\operatorname{R}(T_i), \operatorname{R}(T_j)) + D_a(a_1, b_1), \\
				D_{ERP}(\operatorname{R}(T_i), T_j) + D_a(a_1, g), \\
				D_{ERP}(T_i, \operatorname{R}(T_j)) + D_a(b_1, g)
			\end{array}
			\right\}, & \text{otherwise}
		\end{cases} &
	\end{flalign}
\end{footnotesize}
where $a_1$ and $b_1$ are the first elements of $T_i$ and $T_j$ respectively, and $\operatorname{R}(T)$ denotes the remainder of the trajectory after the first element.

\subsection{Learning-based trajectory similarity measurement methods}
\begin{figure*}[t]
	\centering
	\includegraphics[width=1\linewidth]{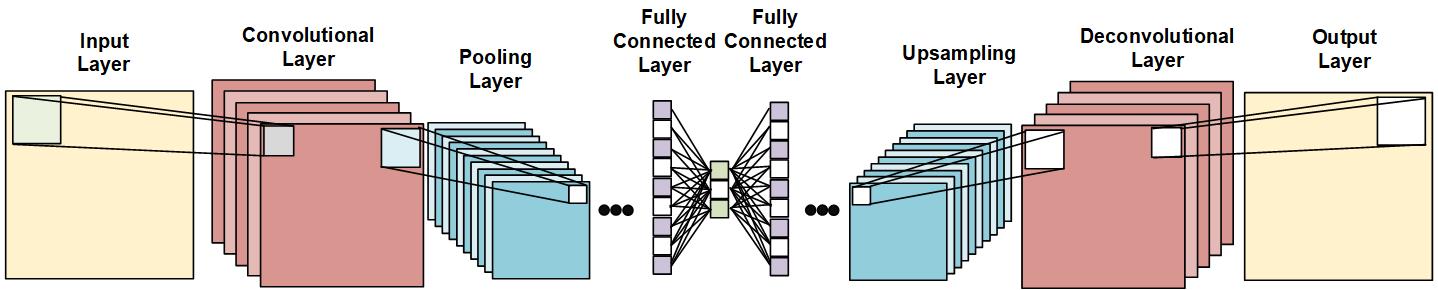}
	\caption{The structure of the Convolutional Auto-encoder model \cite{liang2021unsupervised}.}
	\label{fig:CAE}
\end{figure*}
Traditional methods for measuring trajectory similarity may face three main challenges \cite{yao2018learning}: First, uneven AIS data sampling rates can result in dense or sparse intervals, making it difficult to accurately identify similarities between trajectories. Second, when data points are collected infrequently, such as with geo-tagged tweets or online check-ins, significant gaps in the trajectory data can occur. Third, noisy data, potentially caused by intentional privacy measures or accidental issues like AIS connectivity problems, can further complicate the similarity assessment \cite{taghizadeh2021meaningful}. In recent years, deep representation learning models \cite{wang2024deep} such as word2vec and Glove \cite{yao2017trajectory}, widely recognized in Natural Language Processing (NLP), have significantly impacted tasks like part-of-speech tagging and machine translation. Deep representation learning models address the challenges of varying sentence lengths and inconsistent spatial representations by encoding sentences into fixed-length vectors, ensuring uniformity and comparability across diverse text inputs. Inspired by the success of deep representation learning in the NLP field, numerous researchers have developed learning-based models to measure the similarity of vessel trajectories, leveraging advanced computational techniques to analyze maritime movements. Theoretically, deep representation learning models can overcome various shortcomings of traditional similarity measurement models. It solves multiple problems faced by traditional trajectory similarity measures. In this section, we will delve into several prominent methods that utilize deep representation learning for vessel trajectories similarity measurement. 

\subsubsection{Sequence-to-sequence models}
Sequence-to-sequence (Seq2Seq) models have increasingly found application in the realm of road traffic, particularly for measuring trajectory similarities \cite{taghizadeh2021meaningful}. These models \cite{zhou2023grlstm,chen2023towards}, which originated in natural language processing, have been adeptly adapted to the spatial-temporal domain to encode and decode sequences of locations, capturing the nuances of movement patterns in urban environments \cite{liu2020representation}. For instance, the Traj2Vec model represents a notable advancement in this area, leveraging the sequential nature of trajectory data to encode spatial and temporal attributes effectively. This approach facilitates a deeper understanding of urban land-use types and contributes to sustainable urban development by quantifying residents' spatial trajectories. For example, the Zhang et al.  \cite{zhang2021traj2vec} demonstrates how Traj2Vec can discern the intricate patterns of daily commutes in a bustling metropolis, translating these movements into actionable insights on land-use optimization. This model's ability to process vast amounts of trajectory data illuminates the potential for enhanced urban planning strategies that cater to the evolving needs of urban populations.

The applicability of Seq2Seq models extends beyond the urban context, showing promising results in maritime navigation as well. By adjusting parameters and incorporating domain-specific constraints, such as the vast, open nature of sea routes compared to the constrained urban road networks, these models can capture the complexities of maritime trajectories. This expansion into the maritime domain underscores the versatility of Seq2Seq models in handling various types of trajectory data, offering valuable insights into movement patterns across different environments \cite{gao2020ship}.

The core of Seq2Seq model is an LSTM auto-encoder, which consists of two main components: an encoder and a decoder. The encoder processes the input vessel trajectory $T_i=$ $\left(b_1, b_2, \ldots, b_T\right)$, updating its hidden state at each step according to:
\begin{equation}
	h_t=f_{\mathrm{LSTM}}\left(h_{t-1}, b_t\right)
\end{equation}
where $h_t$ is the hidden state at time $t, b_t$ is the current input from the sub trajectory sequence, and $f_{\mathrm{LSTM}}$ represents the LSTM function.

After the encoder processes the final input $b_T$, the last hidden state $h_T$ serves as a compact representation of the entire input sequence. This representation captures the essential information of the sequence, which is then used by the decoder.

The decoder's task is to reconstruct the original input sequence from the encoded representation. It begins with the final hidden state $h_T$ of the encoder as its initial state and sequentially generates outputs $\left(c_1, c_2, \ldots, c_T\right)$. The decoder's hidden state $h_{d_t}$ is updated using the same form of LSTM function:
\begin{equation}
	h_{d_t}=f_{\mathrm{LSTM}}\left(h_{d_{t-1}}, c_{t-1}, h_T\right)
\end{equation}
where $h_{d_t}$ is the hidden state of the decoder at time $t$, and $c_{t-1}$ is the output generated in the previous step. The objective of this process is the minimization of the difference between the original sequence $T_i$ and the reconstructed sequence $\left(c_1, c_2, \ldots, c_T\right)$, typically employing a mean squared error metric. Through this training, the encoder and decoder are refined to effectively compress and reconstruct the behavior sequences, respectively. 

\subsubsection{Convolutional auto-encoder models}

The Convolutional Auto-encoder (CAE) model \cite{liang2021unsupervised}, initially developed for image processing, is now also used to analyze spatial trajectories. By converting sequences of locations into image-like formats, the CAE model helps to uncover movement patterns useful in urban and maritime mobility. This technique enhances the analysis of spatial patterns and aids in developing efficient planning strategies. In deep learning, representation learning is crucial as it involves encoding useful features from the input data. Among various models, the Autoencoder (AE) and sequence-to-sequence (seq2seq) models are prominent, with the CAE being particularly favored for its ease of implementation and fast training.

In the CAE model, the trajectory data is converted into grayscale images, and the input and output dimensions are consistent. CAE network consists of three parts: convolutional encoder, fully connected layer and convolutional decoder, from input layer to output layer, as shown in Figure \ref{fig:CAE}. The model believes that the vector of the output layer of the fully connected network can be expressed as trajectory data. Calculation of the distance between the vectors can be regarded as the similarity between the trajectories.

In the CAE, the process is divided into two main parts: encoding and decoding. During encoding, the input, which is an image representing a trajectory, is transformed through layers that compress its information. This is done by applying convolution ($Conv$) operations that extract features and max-pooling operations that reduce the size of these features, resulting in a compact representation. The process can be summarized by:
\begin{equation}
	h_{l} = Pool (\operatorname{ReLU}( Conv ( I_{T} )))
\end{equation}
Here, $I_{T}$ represents the input tensor to the encoder. $h_{l}$ represents the output of the encoding phase. $ReLU$ is a type of activation function. In the decoding phase, the goal is to reconstruct the original input from the compressed representation. This is achieved by reversing the encoding steps through deconvolution ($Deconv$) operations that expand the features and unpooling operations ($upSam$) that restore their original dimensions. The decoding process can be described as:
\begin{equation}
	\hat{I_{T}} = Deconv(upSam(h_{l} ))
\end{equation}
where $\hat{I_{T}}$ is the reconstructed output of the decoder. These processes aims to be as close as possible to the original input trajectory $I_{T} $, minimizing the difference between them, which is often measured using a loss function such as mean squared error.

\begin{table*}[]
	\centering
	\caption{Summary of applications related to vessel trajectory clustering.}
	\begin{tabular}{|c|c|c|c|}
		\hline
		Ref.                                   & Applications                                                                & Similarity Measures                            & Clustering method                    \\ \hline
		Pedroche et al. \cite{pedroche2024context}   & Anomaly detection                                                           & DTW                                            & DBSCAN                               \\ \hline
		Xie et al. \cite{xie2023novel}        & Anomaly detection                                                           & DTW                                            & Variational Autoencoder (VAE)        \\ \hline
		Liu et al. \cite{liu2022research}       & Anomaly detection                                                           & Hausdorff distance                             & Immune genetic spectral clustering \\ \hline
		Xie et al. \cite{xie2024anomaly}       & Anomaly detection                                                           & Fast-DTW                                       & DBSCAN                               \\ \hline
		Li et al. \cite{li2024approach}        & Recognize the route pattern & Multi-aspect trajectory similarity measurement & Hierarchical clustering              \\ \hline
		Park et al. \cite{park2021ship}          & Trajectory prediction                                                       & LCSS                                           & Spectral-Clustered                   \\ \hline
		Zhang et al. \cite{zhang2020ais}          & Destination prediction                                                      & Random forest                                  & DBSCAN                               \\ \hline
		Liu et al. \cite{liu2023data}           & Network extraction                                                          & DTW                                            & DBSCAN                               \\ \hline
		Arguedas et al. \cite{arguedas2017maritime} & Network extraction                                                          & Hausdorff distance                             & --                                   \\ \hline
	\end{tabular}\label{tab:appli}
\end{table*}

\section{Unsupervised algorithms for vessel trajectory clustering} \label{sec:Unsupervised}
There are many algorithms for vessel trajectory clustering. Predominantly, the algorithms employed for this purpose fall into two main categories: distance-based and density-based methods \cite{zhang2024predictive,liang2024unsupervised,wei2024adaptive,wang2021ship}. While both approaches offer unique advantages \cite{yang2024harnessing}, this paper concentrates on elucidating the principles and applications of distance-based clustering algorithms, which include the Spectral Clustering (SP), Hierarchical Cluster Analysis (HCA), K-Means, among others. Distance-based clustering algorithms categorize data points into clusters based on the distances between them. This approach is particularly effective for vessel trajectory data, where the spatial proximity and similarity in movement patterns are crucial for cluster formation. The fundamental premise of these algorithms is that trajectories within the same cluster exhibit minimal distances from one another, while those in different clusters are separated by significantly larger distances \cite{shin2024deep}.

\subsection{Spectral Clustering}
SP \cite{tang2021novel} is a versatile and powerful clustering algorithm that has gained prominence for its ability to identify clusters with complex shapes and connectivity patterns, making it particularly useful in vessel trajectory clustering. The computational process of SP can be succinctly described in the following steps:

The first step in SP involves creating a similarity graph from the data. Each data point (in this case, each vessel trajectory) is treated as a node in the graph. Edges between nodes are then established based on some similarity measure, often using the Euclidean distance or Gaussian similarity function (A.K.A the Radial Basis Function, RBF). The choice of similarity measure and the parameters (like the width of the Gaussian kernel) significantly impact the resulting graph structure and, consequently, the clustering outcome.

Once the similarity graph is constructed, the next step is to compute the Laplacian matrix of the graph. The Laplacian is defined as $L = D - A$, where $A$ is the adjacency matrix of the graph (representing edge weights or similarities between nodes) and $D$ is the degree matrix (a diagonal matrix where each element $D[i, i]$ is the sum of the weights of all edges connected to node $i$). The Laplacian matrix captures the graph's connectivity and is key to identifying clusters within the data.

The core of SP lies in the eigenvalue decomposition of the Laplacian matrix. This step involves calculating the eigenvalues and their corresponding eigenvectors. The number of clusters to be identified ($k$) guides the selection of the top $k$ smallest non-zero eigenvalues and their corresponding eigenvectors. These eigenvectors serve as a new representation of the data points in a lower-dimensional space where traditional clustering algorithms (like K-Means) can be more effectively applied.

The rows of the matrix formed by the selected eigenvectors (each row corresponding to a data point in the new representation) are then clustered using a standard algorithm such as K-Means. This step assigns each data point to a cluster based on its position in the lower-dimensional space, effectively segmenting the original data into groups with similar connectivity patterns.

Finally, data points are assigned to clusters based on the clustering results in the transformed lower-dimensional space. Each point is allocated to the same cluster as its corresponding point in the eigenvector-based representation.

\subsection{Hierarchical clustering}
Hierarchical clustering \cite{zhang2024predictive} is a method in data analysis that sequentially divides a dataset layer by layer, ultimately forming a dendrogram, which is a tree-like structure depicting the hierarchical organization of clusters. This methodology can be implemented via two distinct stratagems: the divisive "top-down" approach and the agglomerative "bottom-up" approach. The former initiates the process by considering each data point as an individual cluster and progressively merges the proximally closest clusters in successive iterations until a singular cluster is formed or a predefined cluster count is attained. The agglomerative approach starts by combining all data points into one cluster. It then gradually splits this cluster into smaller ones. This process continues until each data point becomes a separate cluster or a set number of clusters is reached, forming a hierarchical structure.

Within the hierarchical clustering paradigm, the methodology to ascertain the two nearest clusters encompasses several techniques, predominantly:

\subsubsection{Single Linkage} \cite{ros2019hierarchical} This technique defines the distance between two clusters based on the minimum distance between any two points (trajectories) in the different clusters. Despite its simplicity, single linkage is prone to the "chaining phenomenon," where clusters may be linked through a series of proximate points, potentially bridging considerable distances.

\subsubsection{Complete Linkage} \cite{hubert1974approximate} In contrast to single linkage, complete linkage computes the cluster distance as the maximum distance between any two points in the different clusters. This method mitigates the chaining effect but may result in compact, but potentially distant clusters.

\subsubsection{Average Linkage} \cite{charikar2019hierarchical} This method calculates the mean distance between all pairs of points in the two clusters. Average linkage offers a balance between the sensitivity of single linkage to outliers and the potential for over-conservatism in complete linkage.

\subsubsection{Weighted Linkage} \cite{nasibov2011owa} Weighted linkage considers the distance between cluster centroids, thereby providing a measure that accounts for the overall distribution of points within clusters.

\subsubsection{Ward's Method} \cite{murtagh2014ward} Ward's linkage minimizes the total within-cluster variance. At each step, the pair of clusters with the minimum between-cluster distance are merged, ensuring the most compact clusters.

The hierarchical clustering algorithm, with its straightforward principle, facilitates the elucidation of the intrinsic hierarchical relationships within a set of trajectories, offering multi-faceted insights into data structure. Nevertheless, it is noteworthy that hierarchical clustering is computationally intensive and exhibits sensitivity to outliers, which could potentially skew the clustering outcomes.

\section{Applications of Vessel Trajectory Clustering}\label{sec:Applications}
The applications of vessel trajectory clustering are diverse and critical for advancing the maritime industry, ensuring it remains safe, efficient, and environmentally. Vessel trajectory clustering involves grouping similar patterns of vessel movements to understand maritime behavior better and make informed decisions. Derived from the outcomes of vessel trajectory clustering, an array of maritime behavioral characteristics can be discerned, serving a spectrum of purposes. These include the prognostication of vessel trajectories, the detection of aberrant behaviors, and the extraction of maritime traffic networks, among others. Such applications are pivotal in enhancing maritime situational awareness, optimizing navigational efficiencies, and ensuring maritime safety and environmental stewardship. The Table \ref{tab:appli} shows the main application scenarios related to vessel trajectory clustering.

\subsection{Anomaly detection}
Vessel trajectory similarity and trajectory measurement are crucial techniques in the field of anomaly detection, especially in maritime surveillance and safety. These methodologies enable the identification of unusual patterns or behaviors in the movements of vessels and other maritime vessels, which can be indicative of potential threats or irregular activities.
Vessel trajectory similarity involves comparing the paths taken by vessels over time to identify patterns or deviations from common routes. By analyzing historical data and established shipping lanes, this approach can highlight vessels that deviate significantly from expected paths, potentially signaling unauthorized activities, such as smuggling or illegal fishing. Xie et al. \cite{xie2023novel} presents an innovative model that employs a Gaussian Mixture Variational Autoencoder (GMVAE) for detecting anomalies in vessel trajectories. The model is designed to learn the distribution of normal trajectory data and identify deviations that may indicate anomalies. Hu et al. \cite{hu2022intelligent} proposed a Transfer Learning based Trajectory Anomaly Detection (TLTAD) strategy for IoT-empowered Maritime Transportation Systems (IoT-MTS). This approach utilizes a Variational Autoencoder (VAE) to discover potential connections between each dimension of normal trajectories.

\subsection{Trajectory prediction}
The application of vessel trajectory similarity and trajectory measurement extends into the domain of trajectory prediction, playing a pivotal role in enhancing navigational safety and efficiency in maritime operations \cite{zhang2020ais} . In trajectory prediction, the analysis of trajectory similarity helps in modeling typical vessel behaviors and routes based on historical patterns. This insight allows for the anticipation of a vessel's future position with a high degree of accuracy, considering factors such as prevailing traffic conditions, environmental influences, and the vessel's operational characteristics. For example, Alizadeh et al. \cite{alizadeh2021vessel} introduced a novel approach for vessel trajectory prediction, emphasizing the application of DTW to accurately measure trajectory distances, facilitating precise multi-step forecasting and risk assessment for maritime navigation.

\subsection{Other applications}
Vessel trajectory similarity and trajectory clustering techniques find extensive applications in various maritime domains. By analyzing the trajectories of vessels over time, researchers can extract the underlying network of maritime routes \cite{kontopoulos2021distributed,yan2020exploring}, akin to a road map of the sea. This network highlights the most frequented paths, identifying critical maritime corridors and chokepoints. Such insights are invaluable for strategic planning, enhancing the efficiency of shipping routes, and identifying areas that may require additional monitoring or infrastructure development. In the context of vessel route planning, trajectory similarity and measurement techniques enable the optimization of routes for efficiency and safety. By considering historical trajectory data, current sea conditions, and known obstacles, these techniques can suggest optimal paths that minimize travel time and fuel consumption while avoiding hazardous areas. This not only reduces operational costs but also contributes to environmental sustainability by lowering emissions \cite{xin2023maritime}.

\section{Experimental evaluation}\label{sec:Experimental}
In this section, we delve into a comprehensive assessment of the methodologies and technologies applied in clustering vessel trajectories, drawing from the extensive literature review and theoretical groundwork laid in the introduction. This section systematically dissects the evaluation process into several critical components to provide a clear, structured understanding of our experimental approach and the subsequent findings. Specifically, our experiments will evaluate the impact of pre-processing on the clustering of vessel trajectories. Additionally, we will analyze which similarity measurement and clustering algorithms yield superior experimental results, further refining our approach and conclusions.

\subsection{Available software libraries}
\begin{table*}[]
	\centering
	\caption{Related resources.}
	\begin{tabular}{|c|c|c|c|c|c|c|c|}
		\hline
		Algorithm library$\ast$$\ast$ & ELKI   & \begin{tabular}[c]{@{}l@{}}Trajectory\\ distance\end{tabular} & traj-dist & Trajectory Clustering & trajectory-similarity & trajectory-distance-benchmark & trajminer \\ \hline
		Languages                       & Java   & Python                                                        & Python    & Java                  & Java                  & Java   & Python                         \\ \hline
		DTW                           & $\ast$ & $\ast$                                                        &  $\ast$    & $\ast$                & $\ast$                & $\ast$           &              \\ \hline
		LCSS                          & $\ast$ & $\ast$                                                        & $\ast$    & $\ast$                & $\ast$                & $\ast$            & $\ast$       \\ \hline
		EDR                           & $\ast$ & $\ast$                                                        & $\ast$    &                       & $\ast$                & $\ast$            & $\ast$       \\ \hline
		ERP                           & $\ast$ & $\ast$                                                        &  $\ast$    &                       & $\ast$                & $\ast$           &             \\ \hline
		OWD                           &        & $\ast$                                                        &  $\ast$    &                       & $\ast$                & $\ast$           &             \\ \hline
		Fr{\'e}chet                   &        & $\ast$                                                        & $\ast$     &                       & $\ast$                & $\ast$              &          \\ \hline
		Hausdorff                     &        & $\ast$                                                        &  $\ast$   &                       &                       &                     &           \\ \hline
		Others                        & $\ast$ & $\ast$                                                        &           & $\ast$                & $\ast$                & $\ast$             & $\ast$           \\ \hline
	\end{tabular}\label{tab:library}
\end{table*}

\begin{table*}[]
	\centering
	\caption{Correlation library for trajectory clustering and similarity measurement.}
	\begin{tabular}{|l|l|}
		\hline
		Application                                                          & Web Link                                                                                                                     \\ \hline
		\multirow{5}{*}{Introduction of similarity measurement methods}      & ELKI \cite{achtert2009elki} (\url{https://elki-project.github.io/algorithms/distances})                                                                \\ \cline{2-2} 
		& trajectory\_distance \cite{MAIKOL-TRAJ}(\url{https://github.com/maikol-solis/trajectory_distance})                                               \\ \cline{2-2} 
		& traj-dist \cite{bguillouet} (\url{https://github.com/bguillouet/traj-dist})                      \\ \cline{2-2} 
		&       trajectory-similarity \cite{Rena7ssance}   (\url{https://github.com/takhs91/trajectory-similarity})                        \\ \cline{2-2} 
		&\begin{tabular}[c]{@{}l@{}} Trajectory Distances Benchmark \cite{douglasapeixoto} \\ (\url{https://github.com/douglasapeixoto/trajectory-distance-benchmark}) \end{tabular}                       \\ \hline
		\multirow{2}{*}{Similarity measurement method and clustering method} & trajminer\cite{petry2019trajminer}(\url{https://github.com/trajminer/trajminer})                                                                        \\ \cline{2-2} 
		& \begin{tabular}[c]{@{}l@{}}Trajectory Clustering \cite{ivansanchezvera} \\ (\url{https://github.com/ivansanchezvera/TrajectoryClustering})\end{tabular} \\ \hline
		Clustering algorithm library                                         & scikit-learn(\url{https://scikit-learn.org/stable/})                                                                           \\ \hline
	\end{tabular}\label{tabel:libraryweb}
\end{table*}
\begin{figure}[t]
	\centering
	\includegraphics[width=1\linewidth]{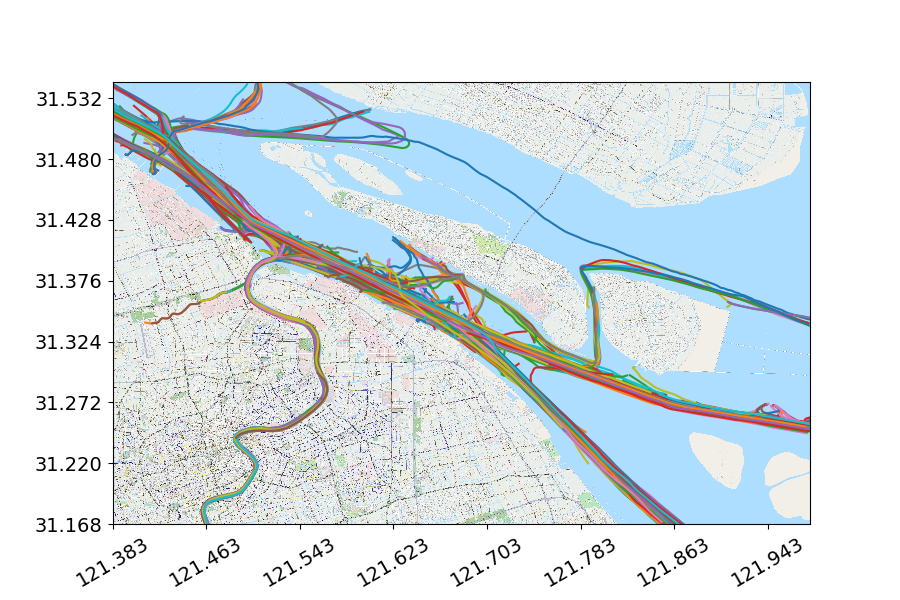}
	\caption{
		AIS Dataset of South Channel of the Yangtze River}
	\label{fig:Shanghai}
\end{figure}
\begin{figure}[t]
	\centering
	\includegraphics[width=1\linewidth]{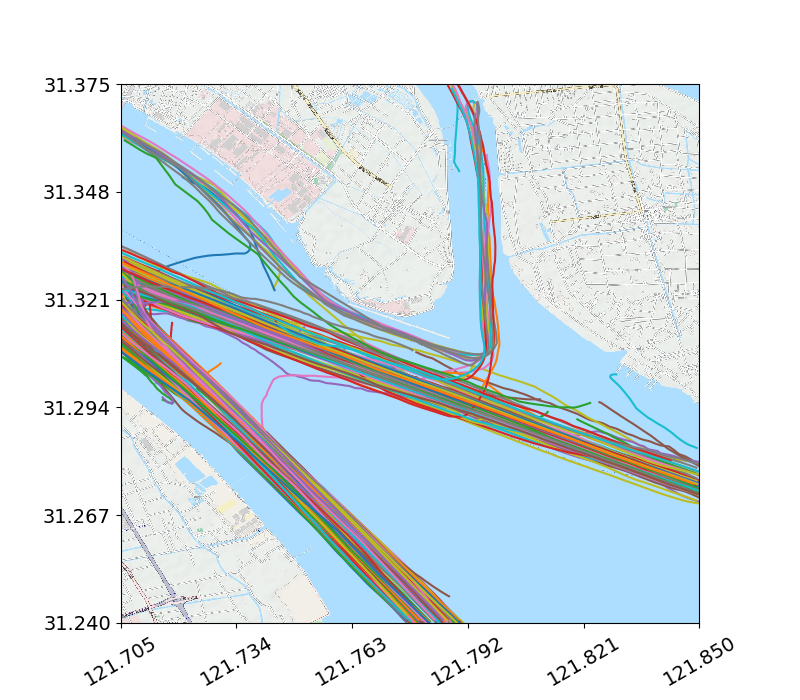}
	\caption{
		A subset of 512 vessel trajectories from Shanghai AIS dataset}
	\label{fig:shsub}
\end{figure}
\begin{table}[]
	\centering
	\caption{
		Efficiency of different trajectory similarity measure algorithms.}
	\begin{tabular}{|c|c|c|}
		\hline
		Distance  & Computation cost          & \begin{tabular}[c]{@{}c@{}}Average computation time \\ (between two trajectories)\end{tabular} \\ \hline
		Hausdorff & $O\left ( n^{2} \right )$ & 0.358996868134   \\ \hline
		Fr{\'e}chet   & $O\left ( n^{2} \right )$ & 0.107457876205   \\ \hline
		OWD       & $O\left ( n^{2} \right )$ & 0.143410921097   \\ \hline
		DTW       & $O\left ( n^{2} \right )$ & 0.110918998718   \\ \hline
		LCSS      & $O\left ( n^{2} \right )$ & 0.105146884918   \\ \hline
		EDR       & $O\left ( n^{2} \right )$ & 0.116922855377   \\ \hline
		ERP       & $O\left ( n^{2} \right )$ & 0.297501802444   \\ \hline
	\end{tabular}\label{tab:Efficiency}
\end{table}
\begin{figure}[t]
	\centering
	\includegraphics[width=1\linewidth]{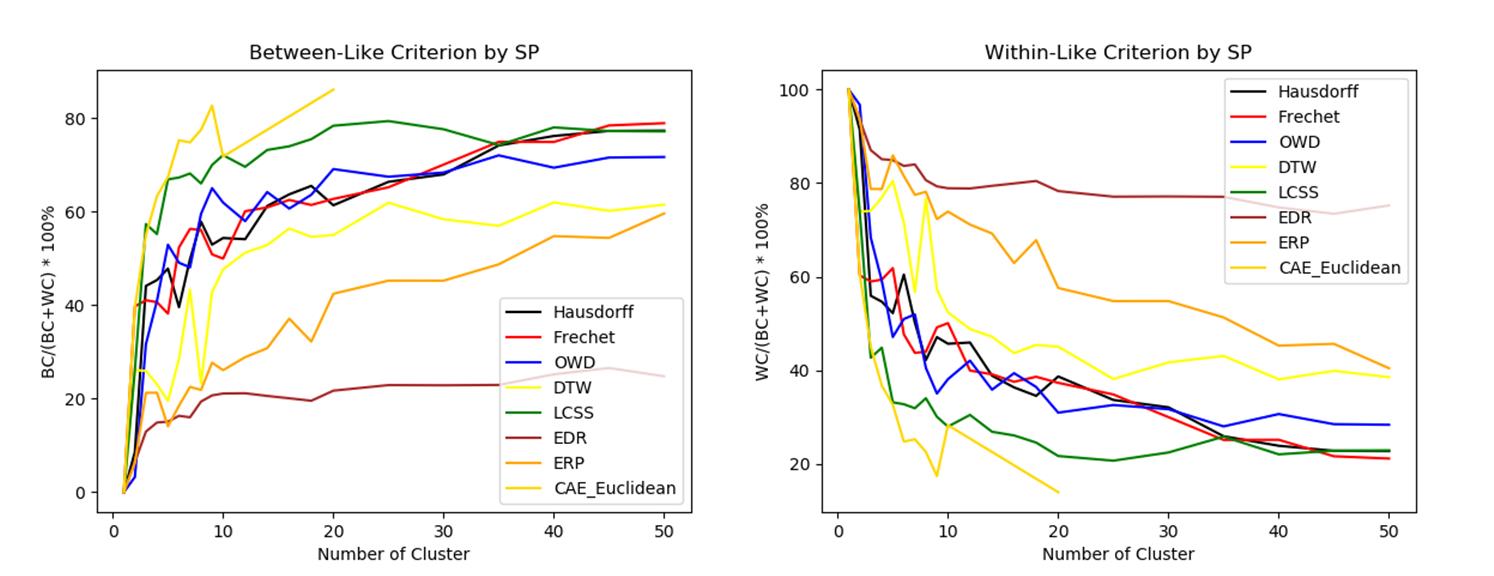}
	\caption{
		SP-based vessel trajectory clustering result.}
	\label{fig:SP}
\end{figure}
\begin{figure}[t]
	\centering
	\includegraphics[width=1\linewidth]{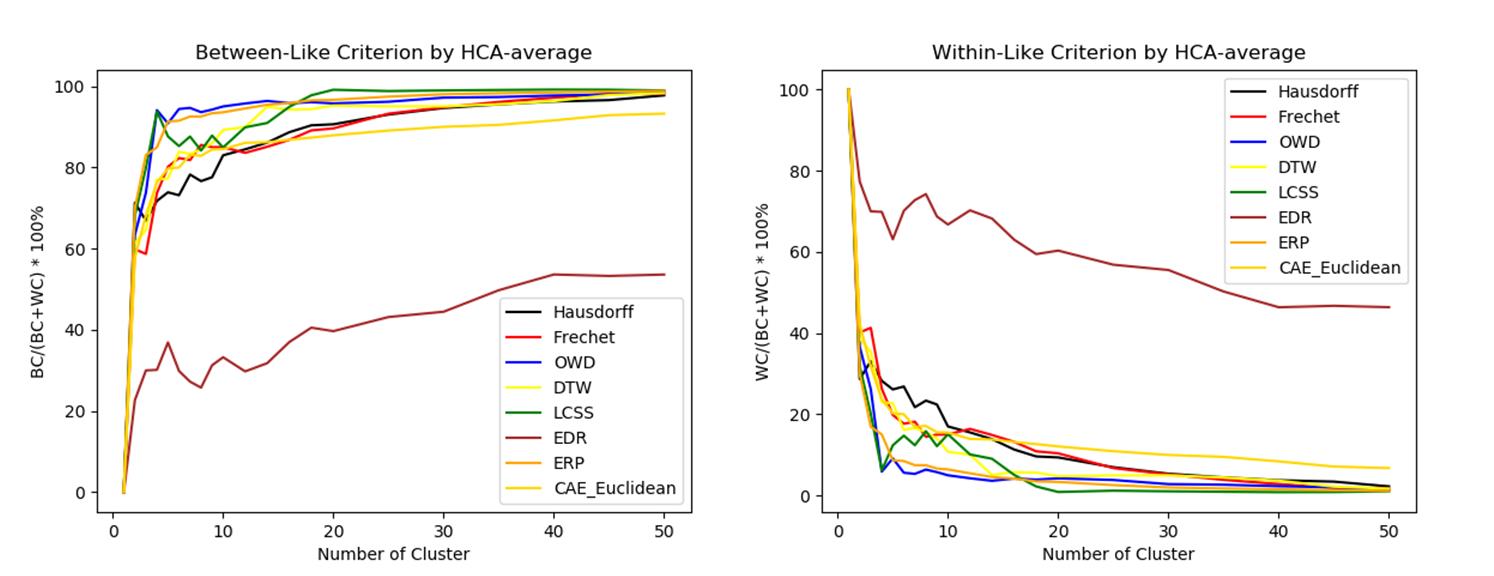}
	\caption{
		HCA-average-based vessel trajectory clustering result.}
	\label{fig:HCA-a}
\end{figure}
\begin{figure}[t]
	\centering
	\includegraphics[width=1\linewidth]{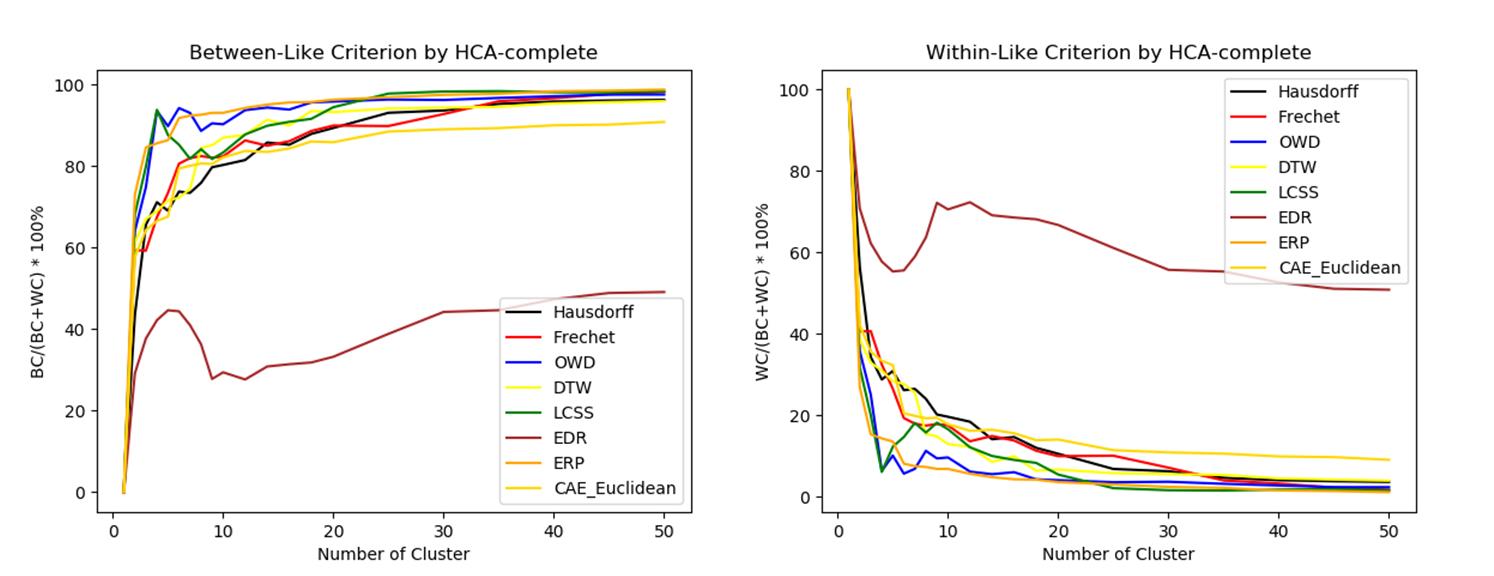}
	\caption{
		HCA-complete-based vessel trajectory clustering result.}
	\label{fig:HCA-c}
\end{figure}
\begin{figure}[t]
	\centering
	\includegraphics[width=1\linewidth]{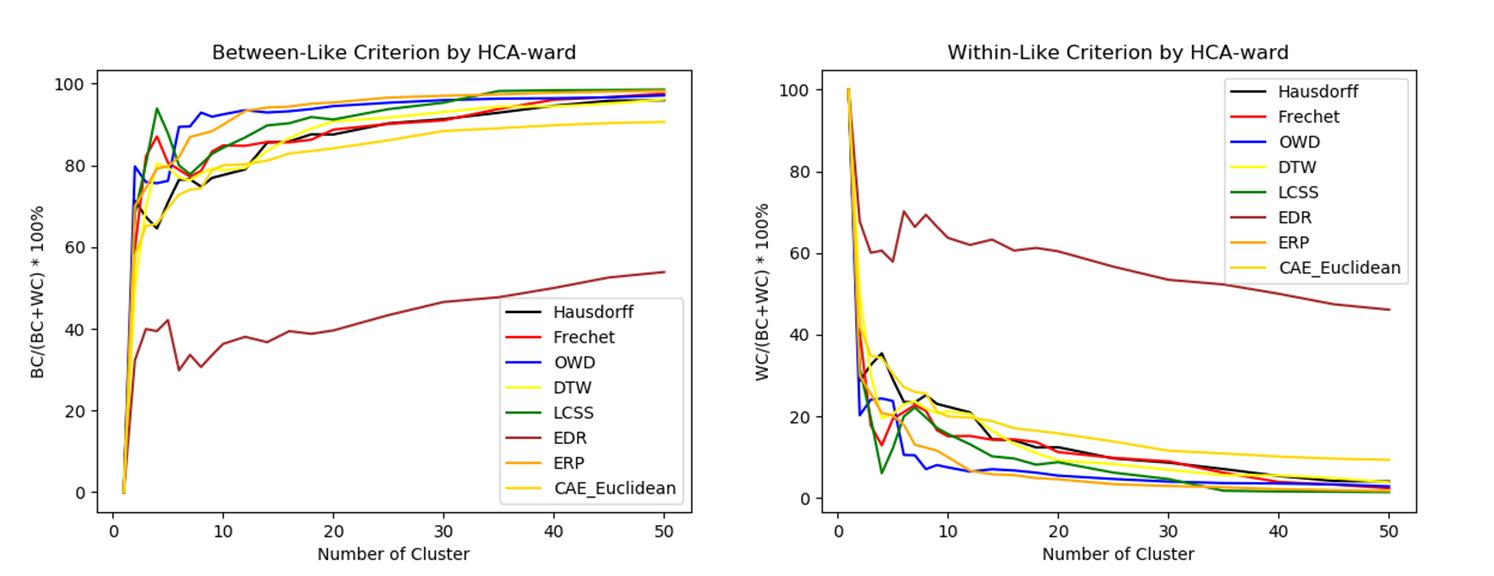}
	\caption{
		HCA-ward-based vessel trajectory clustering result.}
	\label{fig:HCA-w}
\end{figure}
\begin{figure}[t]
	\centering
	\includegraphics[width=1\linewidth]{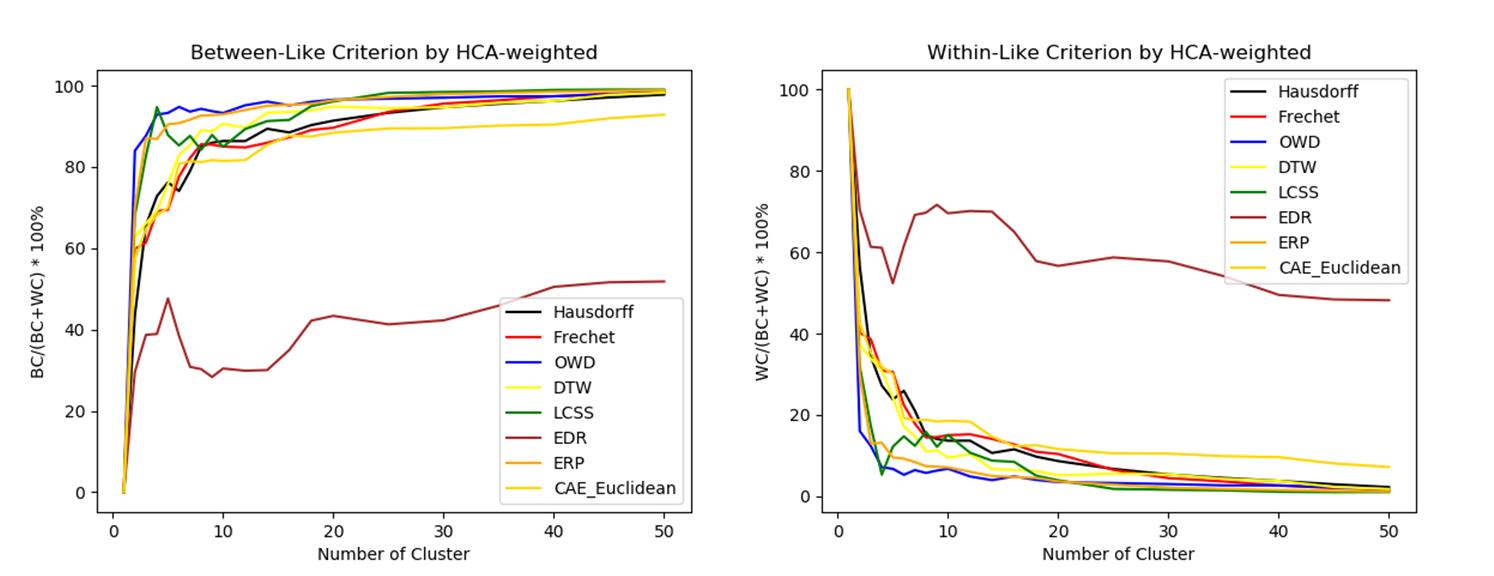}
	\caption{
		HCA-weighted-based vessel trajectory clustering result.}
	\label{fig:HCA-we}
\end{figure}
\begin{figure}[t]
	\centering
	\includegraphics[width=1\linewidth]{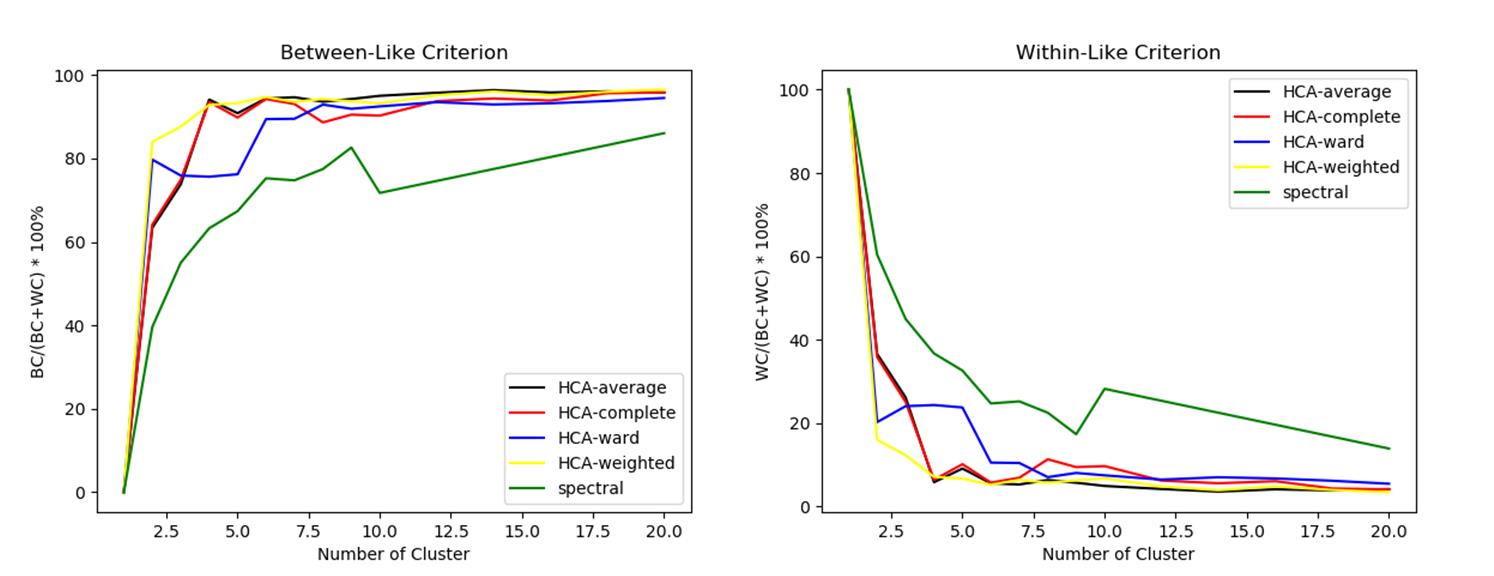}
	\caption{
		OWD-based vessel trajectory clustering results.}
	\label{fig:OWD-V}
\end{figure}

This paper summarizes several libraries for trajectory similarity measurement and clustering, as detailed in Tables \ref{tab:library} and \ref{tabel:libraryweb}. It highlights trajectory similarity measurement methods featured in different libraries, each implementing algorithms in distinct programming languages. We believe that researchers who measure trajectory similarity will find these results very helpful for quickly identifying suitable target algorithms and quickly implementing comparison algorithms, thereby saving time in searching on the Internet. All experiments in this paper were conducted using the trajectory\_distance(\url{https://github.com/maikol-solis/trajectory_distance}) as well as the scikit-learn library.

\subsection{Data Description}

To analyze the effectiveness of various trajectory similarity measurement and clustering algorithms, we utilized data from the AIS base stations, specifically covering the South Channel of the Yangtze River (longitude: 121.383-121.979 , latitude: 31.168-31.546) on July 31, 2017. We simply de-noised the data using our previous studies \cite{liang2021mvffnet} as shown in Fig. \ref{fig:Shanghai}. 
Given the vast number of data points in the vessel trajectories, clustering efficiency is significantly reduced. To compare the differences between various algorithms more swiftly and effectively, we initially focused on a complex subset of the area (longitude 121.705-121.85, latitude 31.24-31.375) for experimental analysis to enhance the efficiency of our comparisons. This subset includes 512 vessel trajectories, as illustrated in Fig. \ref{fig:shsub}.  Subsequently, we expanded our experiments to the complete dataset.

\subsection{Assessment Criteria}
The purpose of vessel trajectory clustering is to separate trajectories with large differences and to aggregate trajectories with similar characteristics \cite{liang2021unsupervised}. In other words, trajectories in the same cluster should be as similar as possible, while the trajectories among clusters should be as distinguishable as possible. Therefore, within group variance and between group variance can be selected as indexes to evaluate the trajectory clustering. 

Before defining the within-group variance and between-group variance, the mean object should be defined. The mean object $T_{Mean}$ for a set of vessel trajectories is given by:
\begin{equation}
	T_{Mean}=\min_{T^{i} \, i\in \left [ 0 ...N\right ]}\left \{ \sum_{j=1,i\neq j}^{N} D(T_{i},T_{j})\right \}
\end{equation}
where $N$ denotes the number of trajectories, $D(\cdot)$ is the similarity measure function.

Then, we can define two criteria to approximate the between and within variance: the $Between-Like$ Criteria (BC) and the $Within-Like$ Criteria (WC). The BC and the WC are defined as:
\begin{equation}
	BC=\sum_{k=1}^{K}D\left ( T_{Mean}^{C_{k}}, T_{Mean}\right )
\end{equation}

\begin{equation}
	WC=\sum_{k=1}^{K}\frac{1}{\left | C_{k} \right |}  \sum_{T^{i}\in C_{k}}D\left ( T_{Mean}^{C_{k}}, T^{i}\right )
\end{equation}
where $C_{1} , . . . , C_{K}$ is a set of clusters of trajectory. 
In general, when evaluating the results of trajectory clustering, a larger BC value is preferable, indicating clearer separation between different clusters. Conversely, a smaller WC value is desirable, as it suggests that the trajectories within each cluster are more similar to each other. In this paper, $BC\setminus (BC+WC)\times 100\%$ and $WC\setminus (BC+WC)\times 100\%$ are used as evaluation indicators.

\subsection{Quantitative analysis}
%

\subsubsection{Computational efficiency analysis}
Table \ref{tab:Efficiency} illustrates the computational efficiency of various trajectory similarity measurement algorithms on the vessel trajectory dataset.
It can be seen that even though the complexity of the trajectory similarity measurement algorithms is the same, the computational efficiency of the algorithms still varies. Notably, the Hausdorff and ERP are calculated with the least efficiency, being approximately three times less efficient than the other algorithms examined.
\begin{figure}[t]
	\centering
	\includegraphics[width=1\linewidth]{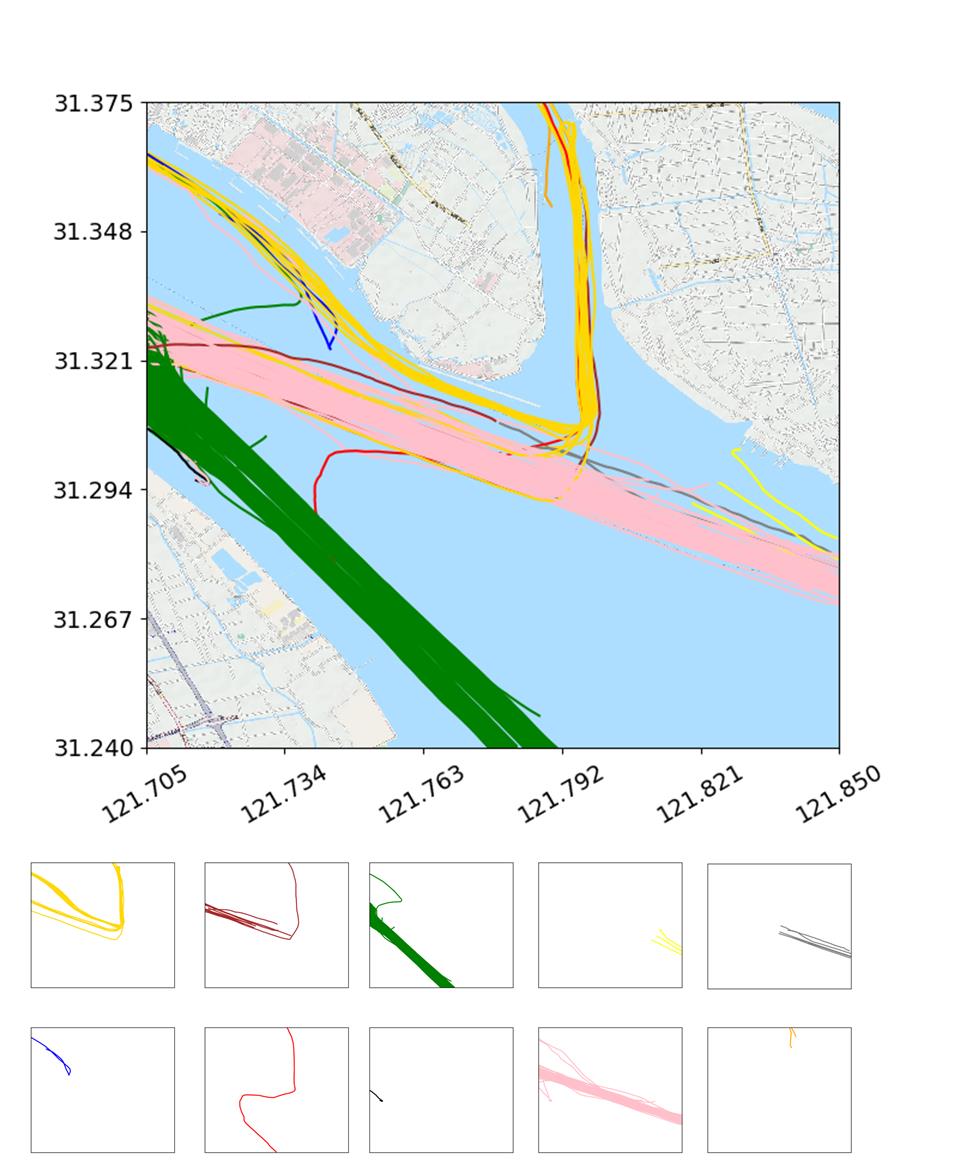}
	\caption{
		Vessel trajectory clustering results based on OWD distance and HCA-weighted.}
	\label{fig:CAE+HCA}
\end{figure}

\subsubsection{Trajectory similarity measurement analysis}
Since there is no ground truth for the similarity of vessel trajectories, the strengths and weaknesses of algorithms for measuring vessel trajectory similarity are generally assessed through vessel trajectories clustering. To analyze the accuracy of different vessel trajectory similarity metric algorithms and clustering algorithms as carefully as possible, this paper exhausts the combinations of relevant algorithms as shown in Figs. \ref{fig:SP}-\ref{fig:HCA-we}. It can be seen from the Fig. \ref{fig:SP} that the CAE and LCSS models achieved better results when using the SP algorithm. In Figs. \ref{fig:HCA-a}-\ref{fig:HCA-we}, it can be seen that OWD achieved better results in vessel trajectory clustering. It can be seen that among all the HCA clustering results, the evaluation index reaches the inflection point when the number of clusters is set to $10$. The inflection point is about $20$ in the SP clustering results. Generally, inflection points are used to determine the optimal number of clusters for trajectory clustering. Before reaching the inflection point, adding more clusters typically leads to significant improvements in evaluation metrics. However, beyond this point, further increasing the number of clusters tends to yield only marginal gains or even causes a deterioration in the metrics, suggesting potential overfitting or an unnecessary increase in complexity.

\subsubsection{Trajectory clustering analysis}
According to the previous analysis, the OWD has shown promising clustering results in similarity measurement evaluations. For comparative analysis, we have chosen hierarchical clustering based on OWD, as depicted in Fig. \ref{fig:OWD-V}. The figure illustrates that the performance of spectral clustering methods falls short compared to hierarchical clustering in terms of both inter-class and intra-class evaluation indicators. Among the hierarchical clustering techniques, the average linkage method (HCA-average) outperforms others. Therefore, we have selected the HCA-average method as the focus of our study. As indicated in Fig. \ref{fig:OWD-V}, there are significant changes in the clustering index when the number of clusters is fewer than 5. Beyond 10 clusters, the index shows negligible variation. Generally, the goal of clustering is to group similar trajectories as closely as possible; hence, the most appropriate number of clusters can be determined based on the intra-class and inter-class similarity indicators.

\subsection{Qualitative evaluation}

%

%
%
\begin{figure}[t]
	\centering
	\includegraphics[width=0.95\linewidth]{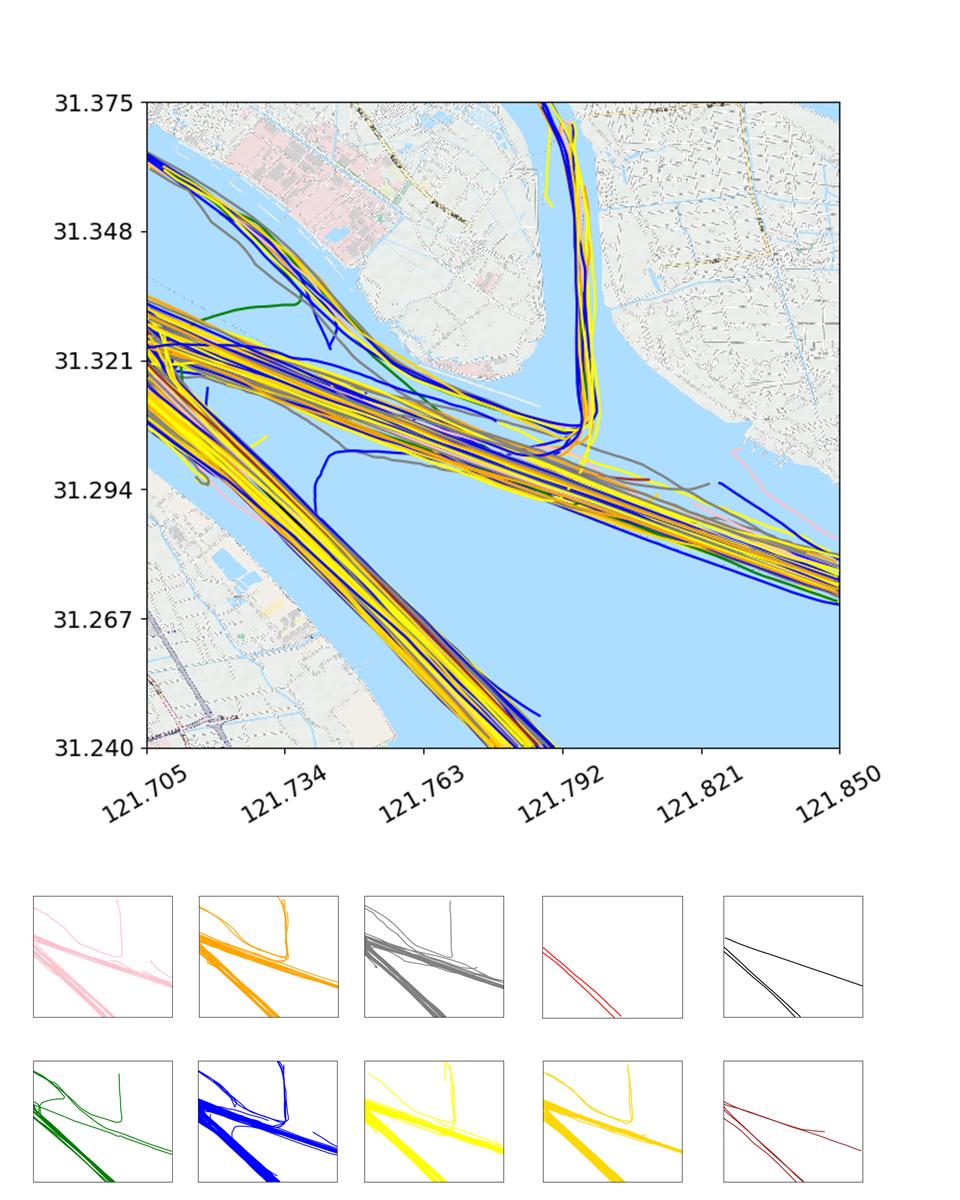}
	\caption{
		Vessel trajectory clustering results based on CAE and spectral clustering.}
	\label{fig:CAE+SP}
\end{figure}
\begin{figure}[t]
	\centering
	\includegraphics[width=1\linewidth]{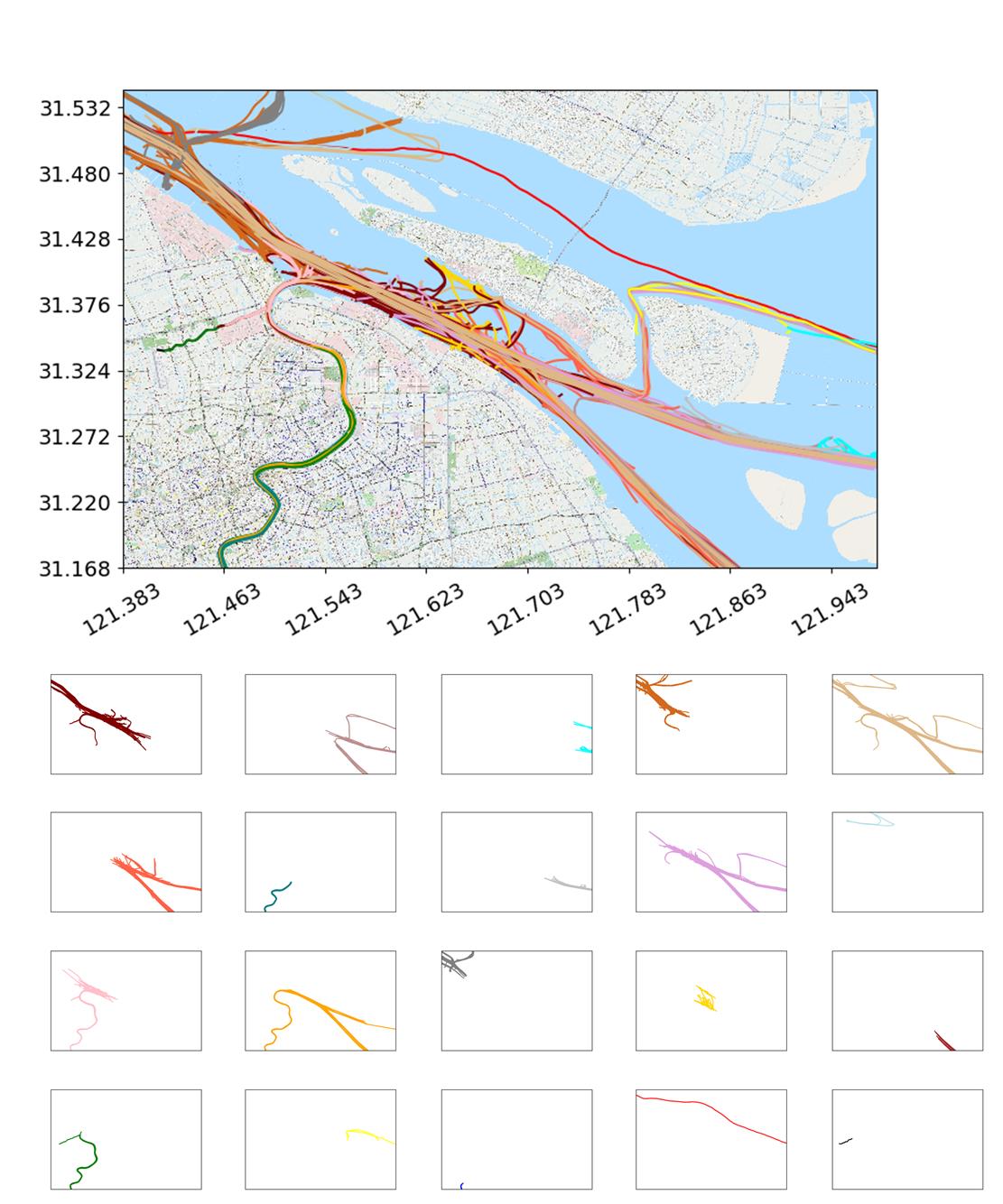}
	\caption{
		Vessel trajectory clustering results on the Shanghai dataset based on OWD+HCA-average.}
	\label{fig:Shangh}
\end{figure}
\begin{figure*}[t]
	\centering
	\includegraphics[width=1\linewidth]{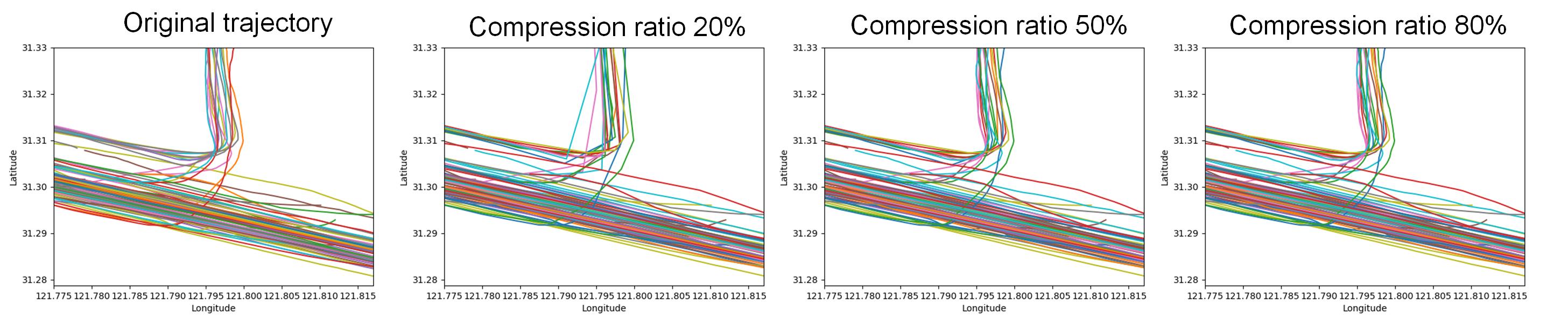}
	\caption{
		Visualization of vessel trajectories with different compression ratios.}
	\label{fig:DPcompression}
\end{figure*}
\begin{figure*}[t]
	\centering
	\includegraphics[width=1\linewidth]{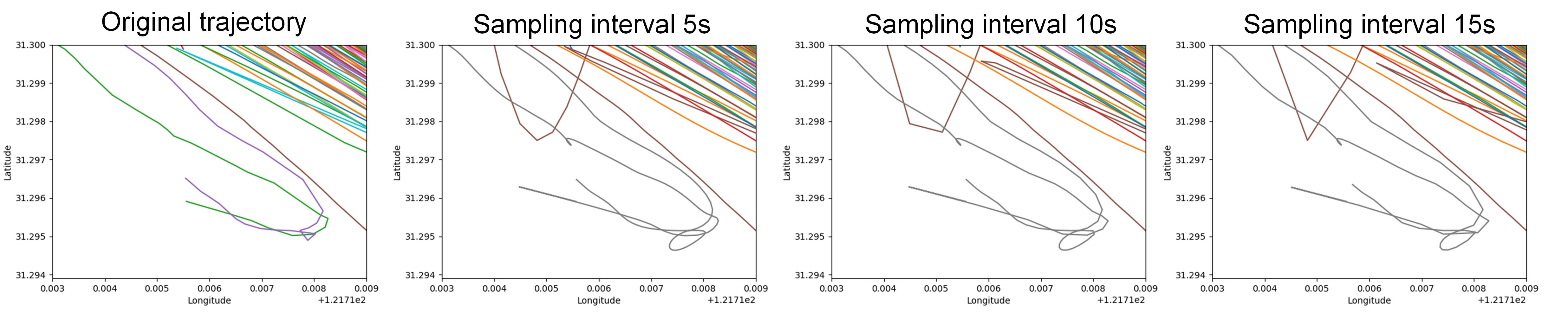}
	\caption{
		Visualization of vessel trajectories with different interpolation intervals.}
	\label{fig:interpolation}
\end{figure*}
\begin{figure}[t]
	\centering
	\includegraphics[width=1\linewidth]{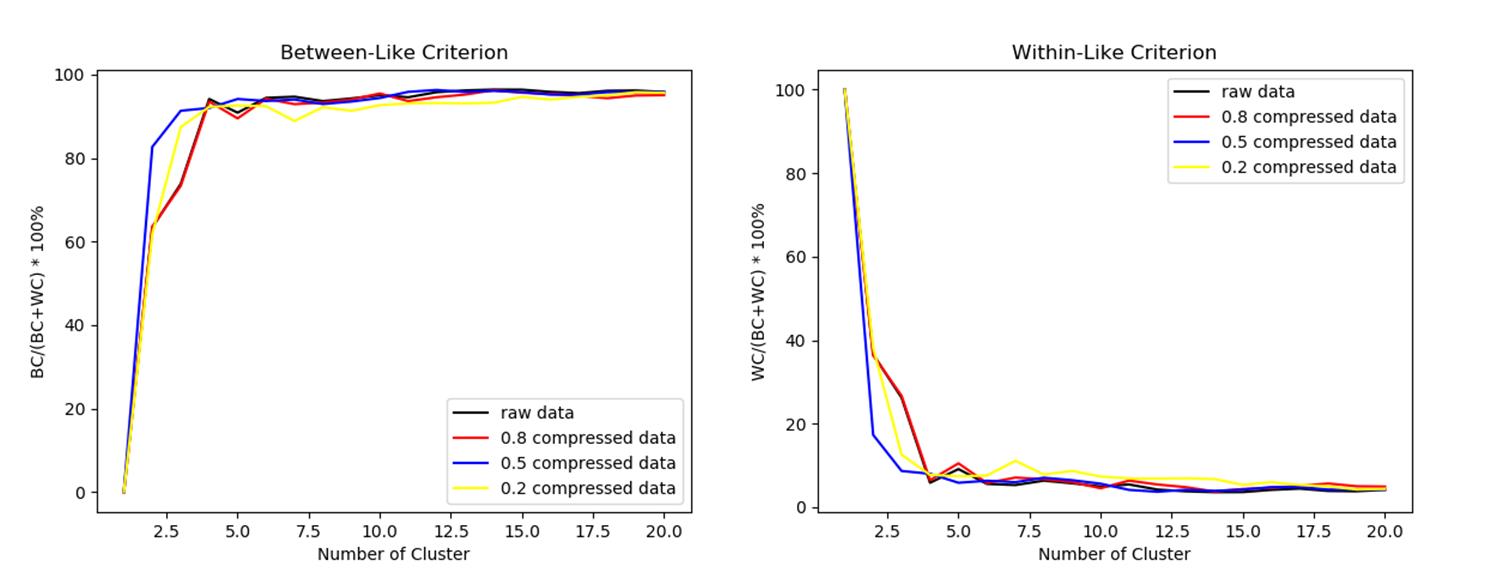}
	\caption{
		Quantitative analysis of vessel trajectory clustering with different compression ratios.}
	\label{fig:DPcompressionQuantitative}
\end{figure}
\begin{figure}[t]
	\centering
	\includegraphics[width=1\linewidth]{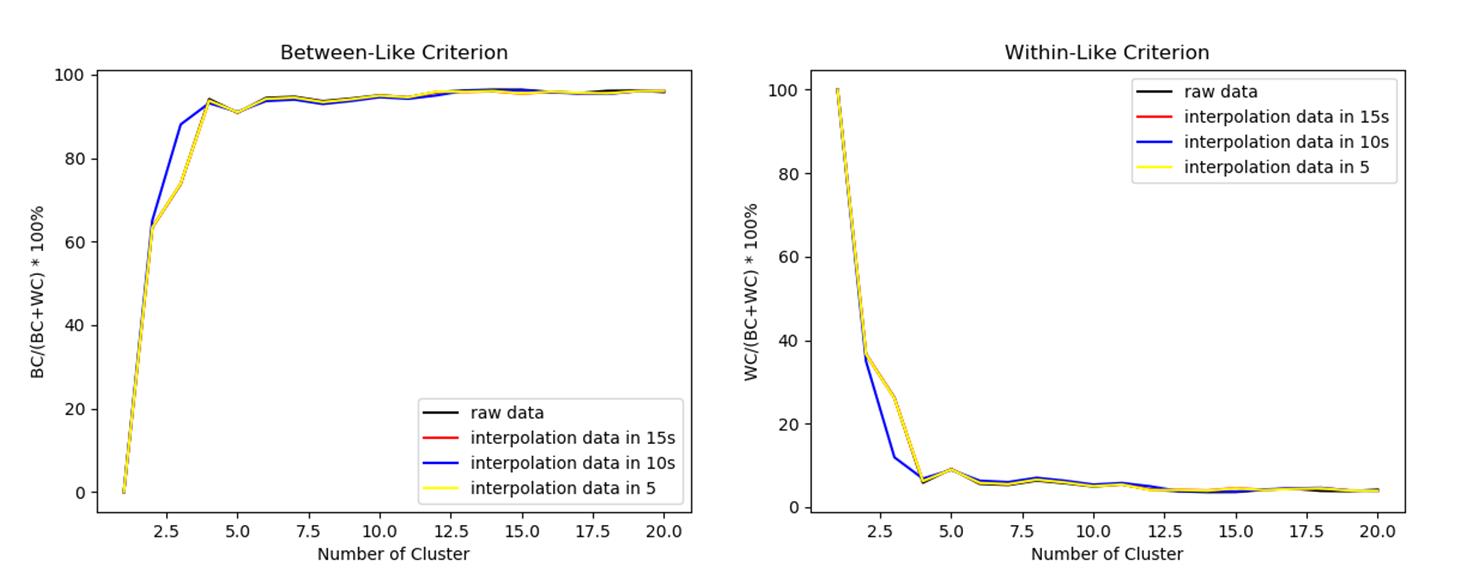}
	\caption{
		Quantitative analysis of vessel trajectory clustering with different interpolation intervals.}
	\label{fig:interpolationQuantitative}
\end{figure}

\begin{table}[]
	\centering
	\caption{The impact of vessel trajectory pre-processing on computational efficiency of trajectory similarity measurement.}
	\begin{tabular}{|cc|ccc|}
		\hline
		\multicolumn{2}{|c|}{\multirow{2}{*}{Number of trajectories}}        & \multicolumn{1}{c|}{10}    & \multicolumn{1}{c|}{50}    & 100    \\ \cline{3-5} 
		\multicolumn{2}{|c|}{}                                               & \multicolumn{3}{c|}{Calculation time (unit: seconds)}             \\ \hline
		\multicolumn{1}{|c|}{\multirow{3}{*}{Interpolation interval}} & 5s   & \multicolumn{1}{c|}{27.18} & \multicolumn{1}{c|}{619.2} & 2231.5 \\ \cline{2-5} 
		\multicolumn{1}{|c|}{}                                        & 10s  & \multicolumn{1}{c|}{12.88} & \multicolumn{1}{c|}{293.3} & 1087.1 \\ \cline{2-5} 
		\multicolumn{1}{|c|}{}                                        & 15s  & \multicolumn{1}{c|}{8.580} & \multicolumn{1}{c|}{195.5} & 740.46 \\ \hline
		\multicolumn{1}{|c|}{\multirow{3}{*}{Compression ratio}}      & 20\% & \multicolumn{1}{c|}{0.085} & \multicolumn{1}{c|}{1.608} & 7.4178 \\ \cline{2-5} 
		\multicolumn{1}{|c|}{}                                        & 50\% & \multicolumn{1}{c|}{0.490} & \multicolumn{1}{c|}{10.27} & 41.538 \\ \cline{2-5} 
		\multicolumn{1}{|c|}{}                                        & 80\% & \multicolumn{1}{c|}{3.350} & \multicolumn{1}{c|}{67.91} & 259.82 \\ \hline
	\end{tabular}\label{tab:efficiency}
\end{table}

According to the quantitative experimental results, OWD+HCA-average and CAE+SP are the two optimal clustering combinations. The number of vessel trajectory clustering clusters were set to $10$. Figs. \ref{fig:CAE+HCA} and \ref{fig:CAE+SP} display the visualization of vessel trajectory clustering outcomes. It is evident that the combination of OWD and HCA-average clustering methods delivers commendable results. From Fig. \ref{fig:CAE+HCA}, it is clear that this algorithm successfully differentiates between vessel trajectory clusters based on their directions and patterns. Conversely, Fig. \ref{fig:CAE+SP} reveals that the clustering outcomes using CAE+SP are less coherent. The algorithm struggles to consistently cluster vessel trajectories that share similar patterns. In addition, clustering experiments were conducted on larger dataset as shown in Fig. \ref{fig:Shangh}. It can be seen that the clustering result of OWD+HCA-average clearly distinguishes the different maritime patterns in the region.

\subsection{Effectiveness of trajectory pre-processing for trajectory clustering.}
What kind of pre-processing is needed before vessel trajectory clustering is controversial. Compression and interpolation are the two main current approaches for vessel trajectory pre-processing. The current mainstream vessel trajectory compression methods and interpolation methods are Douglas–Peucker compression and CSI, respectively. In this paper, different DP compression parameters are chosen to obtain different trajectory compression ratios as shown in Fig. \ref{fig:DPcompression}. It can be seen that there is no significant change in the shape of the vessel trajectory when the vessel trajectory compression ratio is 50\% and 80\%. When the compression ratio is 20\% the vessel trajectory produces a significant jump in the turning region. Fig. \ref{fig:interpolation} shows the results of the visualization of vessel trajectories at different interpolation intervals using CSI. The interpolation results for most vessel trajectories closely resemble the original trajectories. However, some interpolated vessel trajectories exhibit the "Runge Phenomenon," where there are notable deviations from the original trajectories, leading to more pronounced differences.

Fig. \ref{fig:DPcompressionQuantitative} illustrates the results of vessel trajectory clustering for different compression ratios. It can be seen that the evaluation metrics of clustering by compression are improved. The optimal ship trajectory clustering result is obtained when the compression ratio is 50\%. However, as can be shown in Fig. \ref{fig:interpolationQuantitative}, the vessel trajectories clustering results did not change significantly with different interpolation intervals. Tabel \ref{tab:efficiency} shows the impact of vessel trajectory pre-processing on the efficiency of vessel trajectory similarity measurement. It can be seen that compression can significantly improve the computational efficiency of vessel trajectory similarity measures.

\section{Discussion and Conclusion}\label{sec:Discussion}
This survey has meticulously reviewed the utility and ongoing challenges associated with vessel trajectory clustering, a technique essential for improving maritime navigation and safety management. Despite the potential benefits, trajectory clustering remains a complex challenge. This paper focused on recent studies pertaining to distance-based clustering methods for maritime trajectories, providing a critical overview of the field’s development over the past few years.

This comprehensive review covered several core aspects of vessel trajectory clustering: trajectory pre-processing, data, methodologies, applications. Notably, this paper conducts an experimental evaluation of algorithms related to vessel trajectory clustering based on mainstream algorithm libraries.
Our tests focused on the effectiveness of popular distance-based clustering methods. Notably, the OWD method stood out as especially effective. It consistently provided accurate measures of vessel trajectory similarity and showed strong performance throughout our experiments.

Furthermore, we observed that trajectory compression significantly enhances the accuracy and efficiency of similarity measurements. Conversely, our findings indicate that trajectory interpolation, while commonly used, tends to decrease the efficiency of distance-based methods without markedly improving the outcomes of vessel trajectory clustering. This suggests that while interpolation may refine individual trajectory points, it does not necessarily translate to better clustering performance on a broader scale. These conclusions highlight that preprocessing techniques like compression can be more beneficial than interpolation in contexts requiring high efficiency and precision in similarity assessment. However, these findings should not be viewed as universally conclusive; the applicability of vessel trajectory clustering methods can vary greatly depending on the specific scenario and data set involved.

A significant gap in this field is the lack of dedicated datasets and benchmarks for vessel trajectory clustering. This deficiency hampers the ability to standardize and compare the performance of different clustering methods rigorously. Moving forward, it is imperative that future research addresses these gaps by developing comprehensive benchmarks and expanding the datasets available for testing clustering algorithms. By doing so, we can better understand the conditions under which different methods excel and push forward the capabilities of vessel trajectory clustering to meet the diverse needs of maritime traffic management.

\section*{Acknowledgment}

The authors would like to thank Mr. Nanhua Lu at Northwestern Polytechnical University for his help in data analysis.
%
%

%




\end{document}